\documentclass[conference]{IEEEtran}
\IEEEoverridecommandlockouts

\usepackage{cite}
\usepackage{amsmath,amssymb,amsfonts,amsthm}
\usepackage{algorithmic}
\usepackage{graphicx}
\usepackage{textcomp}
\usepackage{xcolor}
\usepackage[hyphens]{url}
\usepackage{fancyhdr}
\usepackage{hyperref}
\usepackage{multirow}
\usepackage{multicol}
\usepackage{soul}
\usepackage{booktabs}
\usepackage[dvipsnames]{xcolor}

\def\BibTeX{{\rm B\kern-.05em{\sc i\kern-.025em b}\kern-.08em T\kern-.1667em\lower.7ex\hbox{E}\kern-.125emX}}

\begin{document}

\pdfpagewidth=8.5in
\pdfpageheight=11in

\newcommand{\iscasubmissionnumber}{3284}

\pagenumbering{arabic}

\title{Fast Cross-Operator Optimization of \\ Attention Dataflow}
\author{
\IEEEauthorblockN{Haodong Chang}
\IEEEauthorblockA{
Texas A\&M University\\
College Station, TX, USA\\
\texttt{haodong@tamu.edu}
}
\and
\IEEEauthorblockN{Hailiang Hu}
\IEEEauthorblockA{
AMD\\
San Jose, CA, USA\\
\texttt{hailiang.hu@amd.com}
}
\and
\IEEEauthorblockN{Zhenrui Wang}
\IEEEauthorblockA{
Texas A\&M University\\
College Station, TX, USA\\
\texttt{zhenruiwang@tamu.edu}
}
\and
\IEEEauthorblockN{Yu Gong}
\IEEEauthorblockA{
Amazon Web Services\\
Santa Clara, CA, USA\\
\texttt{ygkyle@amazon.com}
}
\and
\IEEEauthorblockN{Rongjian Liang}
\IEEEauthorblockA{
NVIDIA\\
Austin, TX, USA\\
\texttt{rliang@nvidia.com}
}
\and
\IEEEauthorblockN{Zhexiang Tang}
\IEEEauthorblockA{
Rutgers University\\
Piscataway, NJ, USA\\
\texttt{zhexiang.tang@rutgers.edu}
}
\and
\IEEEauthorblockN{Bo Yuan}
\IEEEauthorblockA{
Rutgers University\\
Piscataway, NJ, USA\\
\texttt{bo.yuan@soe.rutgers.edu}
}
\and
\IEEEauthorblockN{Jiang Hu}
\IEEEauthorblockA{
Texas A\&M University\\
College Station, TX, USA\\
\texttt{jianghu@tamu.edu}
}
}

\maketitle
\thispagestyle{plain}
\pagestyle{plain}


\begin{abstract}

Attention is a fundamental computational kernel that accounts for the majority of the workload in transformer and LLM computing. Optimizing dataflow is crucial for enhancing both performance and energy efficiency in attention computation. This optimization involves a range of decisions, such as tiling, computation ordering and buffer management, and can be applied at both intra-operator and inter-operator levels, resulting in a highly complex decision space. 

We propose a new approach to cross-operator dataflow optimization. Its centerpiece is an analytical performance model that spans a large decision space and enables matrix-based encoding of multiple candidate solutions. Built on this foundation, a vast number of solutions can be evaluated rapidly, and with the aid of an effective pruning technique, the optimal solution can be identified through exhaustive enumeration. We refer to our method as MMEE (Matrix Multiplication Encoded Enumeration).
The ability to efficiently enumerate a large design space allows MMEE to deliver higher-quality solutions at a substantially faster speed compared to prior approaches.

The MMEE approach is evaluated across various test cases for different accelerator configurations. For energy-driven optimization, MMEE reduces energy consumption by 48\%-50\% and latency by 31\%-69\%, compared to state-of-the-art methods. For latency-driven optimization, MMEE achieves simultaneous reductions of 40\%-50\% in energy consumption and 40\%-69\% in latency, respectively. Additionally, MMEE is $64\times$ to $343\times$ faster than previous works.

\end{abstract}

\section{Introduction} \label{sec:intro}

Attention mechanisms play a central role in transformer-based models, which are prevalent across various application domains, including natural language processing \cite{devlin2019bert, conneau2019cross, raffel2020exploring}, computer vision \cite{dosovitskiy2020image, lin2022cat}, and image generation \cite{zhang2022styleswin, naveen2021transformer}. As models seek to capture correlations across longer contexts, sequence lengths continue to increase \cite{tay2020long, beltagy2020longformer, kitaev2020reformer}. However, the success of attention-based models comes with substantial memory and compute overhead, as the computational complexity of attention scales quadratically with sequence length during prefill and training stages \cite{shen2021efficient, keles2023computational, desislavov2021compute, zadeh2020gobo}.
To address these challenges, numerous techniques have been proposed to improve the efficiency of attention computation on diverse hardware platforms, including CPUs, GPUs, and custom accelerators \cite{dao2022flashattention, shazeer2019fast, ainslie2023gqa, wang2021spatten, shao2023efficient}. Among these platforms, accelerators offer high energy efficiency and reduced latency due to their specialized hardware architectures and flexible dataflow mappings.

\begin{figure}[t]
  \centering
  \includegraphics[width=0.85\linewidth]{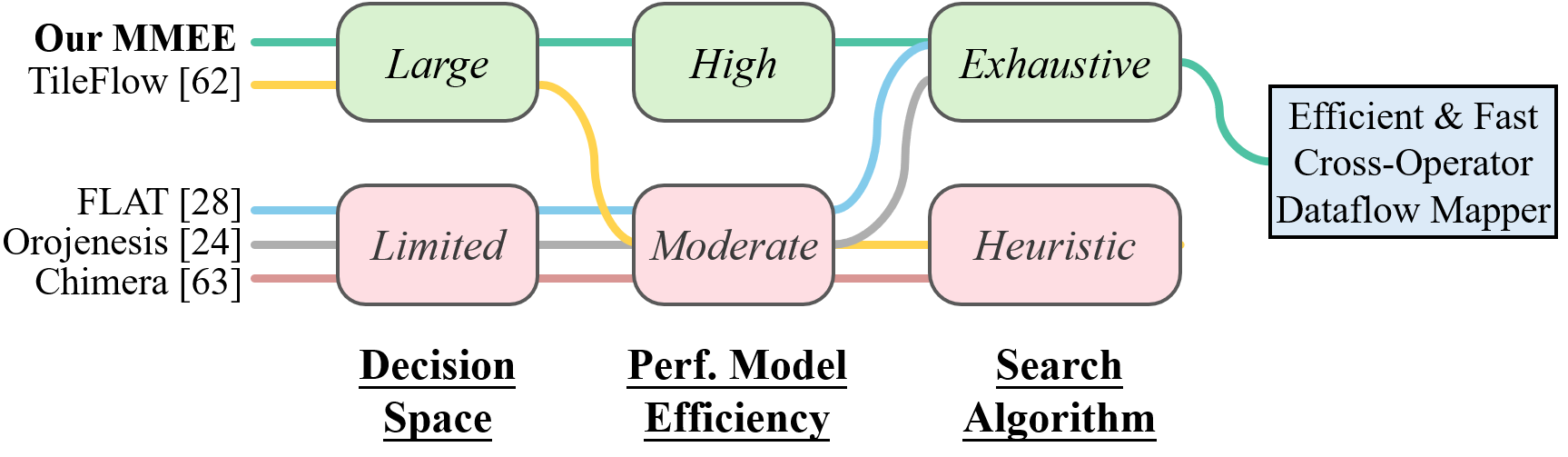}
  \vspace{-1em}
  \caption{Comparison of various cross-operator dataflow mappers.}
  \label{fig:previous_work}
\end{figure}

Dataflow mapping on accelerators dictates how computation and memory resources are utilized spatially and temporally. It not only plays a critical role in determining the efficiency of individual attention workloads, but also provides high-fidelity feedback for accelerator hardware designers. Nevertheless, optimizing dataflow mappings remains a complex and challenging problem~\cite{zheng2023tileflow, nayak2024fusemax, zheng2023chimera}.
Broadly, dataflow optimization can be categorized into intra-operator and inter-operator mappings. Intra-operator optimization focuses on improving the execution of individual operators~\cite{liang2022deep, kao2020gamma, fu2021auto, parashar2019timeloop}, such as matrix multiplication or convolution. Due to data dependencies across operators, inter-operator optimization, such as operator fusion, offers substantial additional opportunities for improving the overall efficiency of attention computation.

In addition to the quality of dataflow mapping solutions, optimization speed is critical in practice. In AI accelerator design, dataflow mapping serves as a core component that is repeatedly invoked when evaluating various hardware architectures and model workloads. Similarly, in AI compilers, the dataflow mapper may be frequently executed to accommodate model modifications. Despite its practical significance, the runtime efficiency of cross-operator dataflow optimization has received limited attention in prior research. \label{para:more_motivation}

How can we obtain high-quality cross-operator dataflow solutions at high speed? We identify three key requirements:
(1) {\em A sufficiently large decision space.} If the decision space is too restricted or omits important decision options, opportunities for discovering better solutions will be lost.
(2) {\em A fast and accurate performance model.} The model used to estimate latency and energy must be computationally efficient so that a large number of candidate solutions can be evaluated within a reasonable runtime.
(3) {\em A systematic and efficient search algorithm.} Without systematic exploration, high-quality solutions may be overlooked; without efficiency, good solutions cannot be found in practical time. 
If any of these requirement is not satisfied, a bottleneck is formed to hinder quick finding of high-quality solutions. 

In light of these requirements, one can see that existing approaches leave room for improvement. We analyze several state-of-the-art methods and summarize their characteristics with respect to these requirements in Fig.~\ref{fig:previous_work}.
Regarding the decision space, both FLAT~\cite{kao2023flat} and Chimera~\cite{zheng2023chimera} omit buffer management, while Orojenesis~\cite{huang2024mind} relies on a limited set of templates to constrain the search space. Consequently, even though FLAT and Chimera perform exhaustive enumeration within their respective spaces, they inherently miss certain opportunities for improvement due to incomplete decision coverage. 
All four approaches, FLAT, Chimera, Orojenesis, and TileFlow~\cite{zheng2023tileflow}, require parsing of decision scenarios or templates before applying analytical formulas. For instance, TileFlow employs a tree representation, and its model evaluation involves tree traversal. Such parsing significantly slows down performance estimation.
TileFlow's search procedure further relies on genetic algorithms and Monte Carlo Tree Search, both of which are randomized heuristics. As a result, they may miss high-quality solutions that exhaustive enumeration would otherwise uncover.

Our focus is on encoder-style and prefill-style attention workloads, where queries are matrix-form and attention remains both compute- and memory-intensive, such as the prefill stage in PD-disaggregated LLM inference systems \cite{zhang2025spad, zhong2024distserve}. Accordingly, our goal is to optimize accelerator-level dataflow rather than system-level scheduling for serving workloads.

First, we capture a broad decision space, including fusion, tiling, computation ordering, buffer management, and recomputation. Second, we introduce an analytical performance model free of ``if--else'' branches, obtained through a significant extension of an intra-operator model~\cite{liang2022deep}. This model enables encoding a large number of candidate solutions in matrix form and evaluating them efficiently via matrix multiplication. Third, we propose an exhaustive enumeration scheme with pruning that explores a large decision space without sacrificing optimality.
We refer to our approach as \emph{Matrix Multiplication Encoded Enumeration} (MMEE). As illustrated in Fig.~\ref{fig:previous_work}, compared to prior state-of-the-art methods, our MMEE is the only one that excels among all three key requirements, leading to high-quality solutions at high speed. 

The contributions of our work are summarized below.
\begin{itemize}
   \item We demonstrate that, despite the imperfectly nested structure of fusion dataflows, a pseudo nested loop representation provides a systematic and concise abstraction that fully captures the decision space defined by tiling, computation ordering, and buffer management.
   \item We develop the first analytical model without traversal procedures, to the best of our knowledge, for estimating buffer size requirements, DRAM accesses, energy consumption and latency for fused attention dataflows. 
    \item We enable rapid and optimal dataflow search by combining matrix-based encoding with an effective pruning strategy. This allows exhaustive evaluation of a large design space and ensures the optimal solution can be identified without approximation.  
\end{itemize}

We evaluate MMEE across various workloads on two accelerators. Compared to state-of-the-art methods, the energy-driven version of MMEE reduces energy consumption by 48\%-50\% and latency by 31\%-69\%. For latency-driven optimization, MMEE reduces energy consumption by 40\%–50\% and latency by 40\%–69\%. Notably, our optimization process is $64\times$ to $343\times$ faster than prior approaches.

\section{Background}

\subsection{Attention Computation in Prefill and Training} \label{sec:attention computation background}

\begin{figure}[t]
  \centering
  \includegraphics[width=\linewidth]{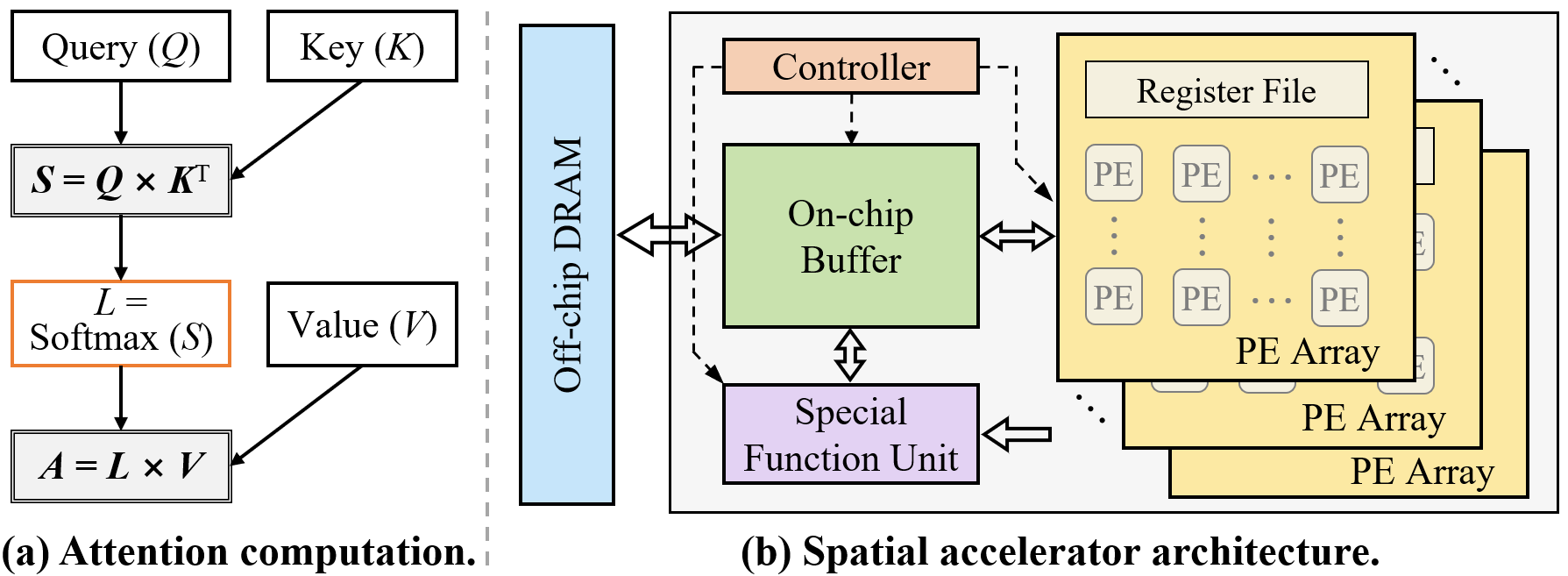}
  \vspace{-2em}
  \caption{Attention computation and accelerator architecture.}
  \label{fig:acc}
\end{figure}

The main operations of attention computation are illustrated in Fig. \ref{fig:acc}(a), where $Q$, $K$, $V$, $S$, $L$, and $A$ represent matrices. Each attention head performs the same computation independently. The actual computational load is quite large due to: (1) matrices $S$ and $L$ scale quadratically with sequence length; and (2) frequent invocation of the attention kernel, often exceeding 100 times, due to multiple heads and transformer layers. As shown in \cite{kao2023flat}, the primary benefit of operator fusion arises from combining the matrix multiplications for $S$ and $A$. However, the presence of softmax operations in between makes the fusion, especially tiled fusion, challenging.

Following FLAT \cite{kao2023flat} and TileFlow \cite{zheng2023tileflow}, we focus on attention in encoder, prefill, and training stages, where queries span the sequence dimension and incur quadratic complexity \cite{keles2023computational, kao2023flat}, rather than degenerating to vectors as in decoding. Unlike decoding workloads with vector-form queries and linear complexity \cite{pan2025fasttree, ho2024block}, these matrix-form attention problems exhibit higher arithmetic intensity and require careful co-optimization of computation and memory access, leading to either memory- or compute-bound behavior (Figs.~\ref{fig:compare with Tileflow on edge} and \ref{fig:compare with Tileflow on cloud}).


\subsection{Accelerator Architecture} \label{sec:acc arch}

Spatial accelerators, also referred to as tiled accelerators, have become increasingly prevalent in accelerator design \cite{zheng2023tileflow, cai2023inter, zhou2024ml, jouppi2017datacenter, chang2023multifuse}. As illustrated in Fig.~\ref{fig:acc}(b), a typical spatial accelerator comprises multiple PE arrays, an on-chip buffer, a special function unit (SFU), and a controller \cite{kao2023flat, chen2020communication, cai2022optimus}. Each PE array, e.g., systolic array, comprises a 2D array of processing elements (PEs) and associated register files, supporting operations such as matrix multiplication and convolution. The on-chip buffer sits between the off-chip DRAM and register files. SFU is responsible for executing specialized functions, such as the softmax operation in attention and layer normalization following feed-forward neural network layers \cite{kao2023flat, shao2023efficient}. These components are orchestrated by a dedicated controller to ensure efficient execution.

\section{Decision Elements in Fusion Dataflow}

In this section, we define fusion and three decision elements: tiling, computation ordering and buffer management. These elements critically influence fusion dataflow in terms of buffer capacity requirements, DRAM traffic, and compute utilization. For simplicity, we temporarily exclude the softmax operator and focus on the fusion of two matrix multiplications.

\subsection{Fusion Dataflow}

\begin{figure}[t]
  \centering
  \includegraphics[width=0.97\linewidth]{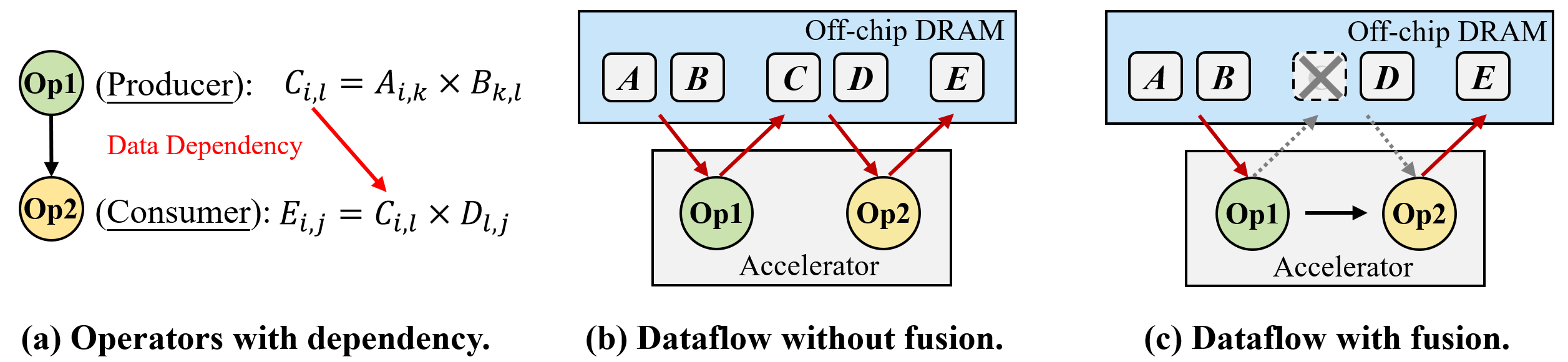}
  \vspace{-0.8em}
  \caption{Fusion dataflow keeps intermediate results ($C$) on-chip, avoiding off-chip DRAM accesses.}
  \label{fig:new_fusion}
\end{figure}

In Fig. \ref{fig:new_fusion}(a), the first multiplication acts as the producer operator, generating the intermediate matrix $C$, while the second serves as the consumer operator, which takes $C$ as the input. Thus, there is data dependency between the consumer and producer operators. 
Fusion represents a dataflow in which the consumer reuses intermediate results without incurring off-chip memory accesses \cite{huang2024mind, kao2023flat, alwani2016fused}, as illustrated in Fig.~\ref{fig:new_fusion}(b) and (c). We abstract three key decision elements that govern such fusion dataflows, detailed in the following sections.

\subsection{Tiling}

\begin{figure}[t]
  \centering
  \includegraphics[width=1.0\linewidth]{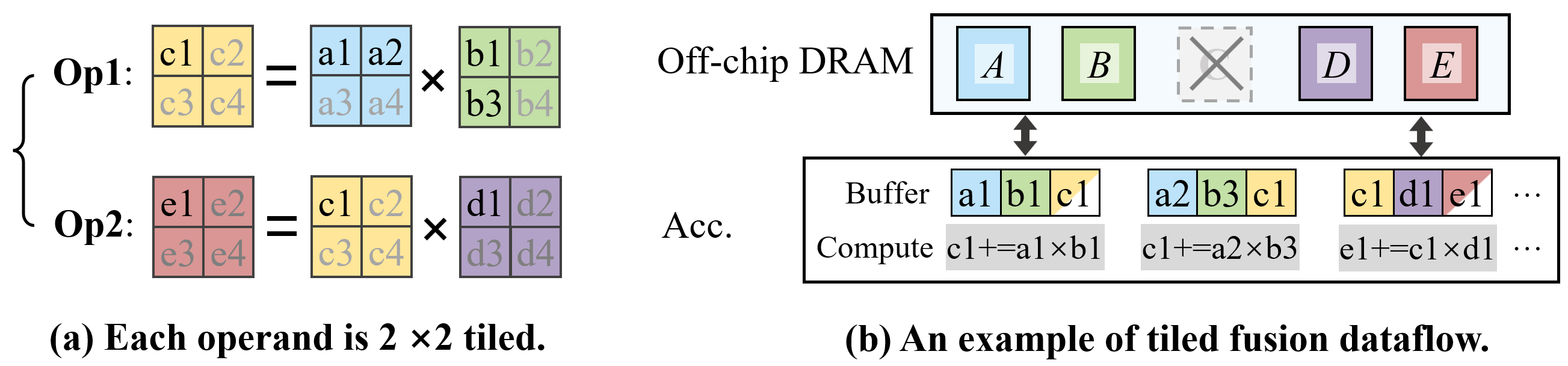}
  \vspace{-2em}
  \caption{Tiling and tiled fusion. (a) Each operator is 2×2 tiled. (b) One tile of intermediate result ($c1$) is kept on-chip instead of $C$, reducing buffer use from 4 tiles to 1. Half-filled square denotes partial sums.}
  \label{fig:new_tiled_fusion}
\end{figure}

Tiling partitions operands into multiple tiles, a common technique to balance the buffer capacity and DRAM access overhead \cite{liang2022deep, kao2023flat, mei2023defines, parashar2019timeloop}.
In fusion dataflow, where intermediate results are often too large to fit entirely in on-chip buffers, tiling can be extended to intermediate outputs, as shown in Fig. \ref{fig:new_tiled_fusion}. Compared to naive fusion, which keeps the entire intermediate matrix $C$ (four tiles) in the buffer, tiled fusion only allocates one tile-sized buffer space for $c1$.

Tiled fusion is particularly critical for attention computations, where the intermediate matrix scales quadratically with the sequence length. In Fig. \ref{fig:new_tiled_fusion}(b), applying tiled fusion reduces the buffer demand for storing intermediate results from four tiles to one. Half-filled square denotes a tile with partial sums. 

\begin{figure}[t]
  \centering
  \includegraphics[width=0.97\linewidth]{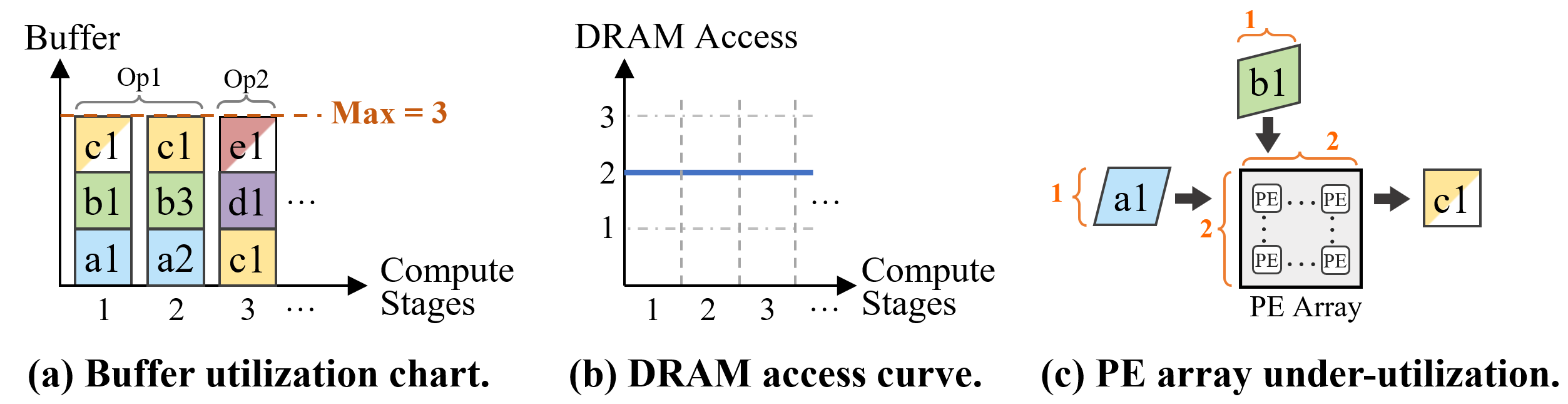}
  \vspace{-0.8em}
  \caption{The impact of tiling. Each compute stage represents the multiplication of a pair of tiles. (a) Buffer utilization and (b) DRAM access curves show buffer usage and off-chip traffic with tiling. (c) When the tile size is smaller than the PE array, the PE array is under-utilized.}
  \label{fig:new_tiling_element}
\end{figure}

Fig. \ref{fig:new_tiling_element} further illustrates the impact of tiling, and the fusion dataflow follows Fig. \ref{fig:new_tiled_fusion}.
The buffer utilization chart (Fig. \ref{fig:new_tiling_element}(a)) and DRAM access curve (Fig. \ref{fig:new_tiling_element}(b)) visualize the dataflow and how dataflow influences on-chip buffer and off-chip traffic. In the buffer utilization chart, the horizontal axis denotes computation stages, each corresponding to the multiplication of a tile pair, while the vertical axis represents buffer allocation. In the DRAM access curve, the vertical axis indicates the number of tiles fetched from DRAM at each stage. If the required tiles are not in the on-chip buffer, DRAM accesses occur. 

As shown in Fig. \ref{fig:new_tiling_element}(a), tiling significantly reduces the overall buffer requirement to three tiles in this example. However, overly aggressive tiling may produce tile sizes smaller than the PE array, as illustrated in Fig. \ref{fig:new_tiling_element}(c), leading to PE under-utilization. This under-utilization may increase latency in compute-bound scenarios, as discussed in Fig. \ref{fig:compute utilization}.

\subsection{Computation Ordering} \label{sec:new_compute_order}

\begin{figure}[t]
  \centering
  \includegraphics[width=0.91\linewidth]{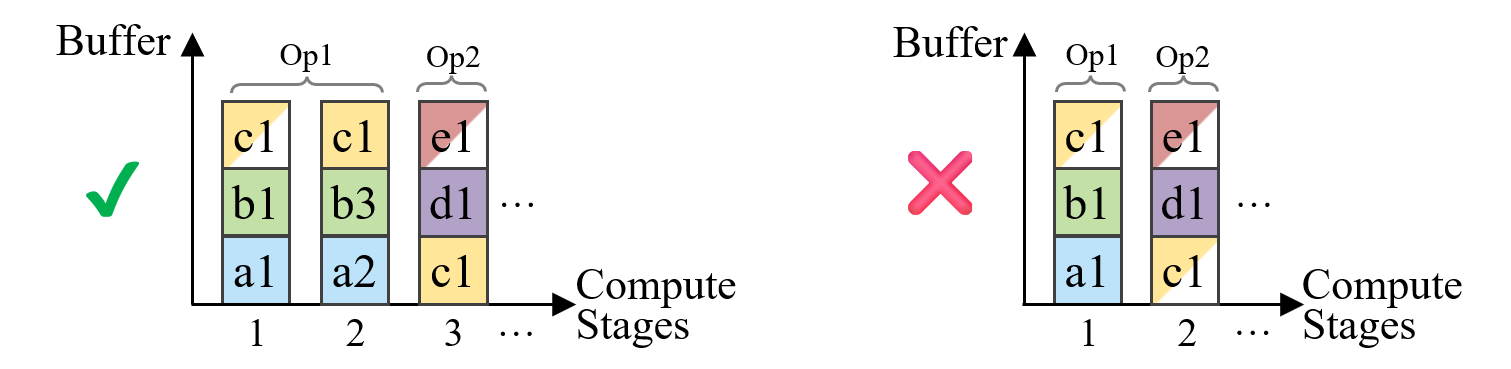}
  \vspace{-1em}
  \caption{The computation ordering on the left is valid, while the right case is invalid because attention fusion requires $c1$ with complete sums.}
  \label{fig:new_compute_order}
\end{figure}

Unlike intra-operator dataflow, the computation ordering for attention fusion is more complex since it interacts with data dependencies, the softmax operator, and recomputation.
Fig. \ref{fig:new_compute_order} compares two fusion dataflows, where only the left dataflow is valid. The right dataflow violates one key constraint required for correct attention fusion, summarized below.

\noindent
{\bf Constraint – No Psum Propagation.}
Intermediate tiles containing partial sums (psums) cannot be forwarded from the producer to the consumer. In the valid dataflow (left), the partial sum of $c1$ produced at stage~\textcircled{1} cannot be consumed at stage~\textcircled{3}; only the fully accumulated tile after stage~\textcircled{2} is allowed. 
However, the right dataflow violates this constraint by consuming $c1$ at stage~\textcircled{2}. This restriction follows FlashAttention \cite{dao2023flashattention, milakov2018online}, where each intermediate tile must be fully accumulated before the online softmax.
By enforcing this constraint, we can systematically generate any valid attention fusion dataflow without violating producer–consumer dependencies, while ensuring functional correctness. 

\begin{figure}[t]
  \centering
  \includegraphics[width=1.0\linewidth]{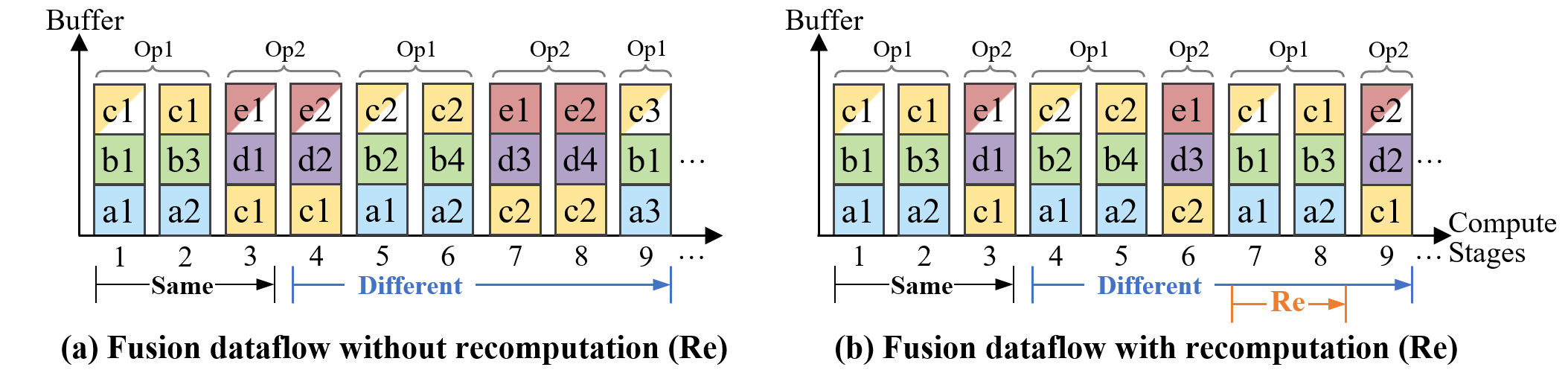}
  \vspace{-2em}
  \caption{Recomputation is included in the computation-ordering space. The two dataflows diverge at stage~\textcircled{3}, and stages~\textcircled{7}–\textcircled{8} in (b) perform recomputation.}
  \label{fig:new_recompute}
\end{figure}

\noindent
{\bf Recomputation.} 
Recomputation enables more flexible dataflows and shorter reuse distances, which in turn reduces DRAM access.
Our framework naturally incorporates recomputation.
Recomputation occurs when intermediate results are evicted and later regenerated from the producer. In Fig.~\ref{fig:new_recompute}(a), stages~\textcircled{1}–\textcircled{3} match those in Fig.~\ref{fig:new_recompute}(b). The divergence begins at stage~\textcircled{4}. The dataflow without recomputation completely consumes $c1$ by performing $c1 \times d2$, while the dataflow with recomputation switches back to the producer (Op1). As a result, $c1$ must be regenerated later. At stages~\textcircled{7}–\textcircled{8}, $c1$ is recomputed, mirroring stages~\textcircled{1}–\textcircled{2}.

\subsection{Buffer Management} \label{sec:new_buffer_manage}

This decision element concerns how on-chip buffer space is allocated for operands. In intra-operator dataflow mapping, this may be expressed by selecting a buffering level \cite{stoutchinin2019optimally, liang2022deep}. However, cross-operator dataflow introduces additional complexity. That is, while the accelerator is executing the producer operator, additional buffer space may be allocated 
to the consumer in order to reduce future DRAM traffic by avoiding repeated loads of consumer operands.

\begin{figure}[t]
  \centering
  \includegraphics[width=1.0\linewidth]{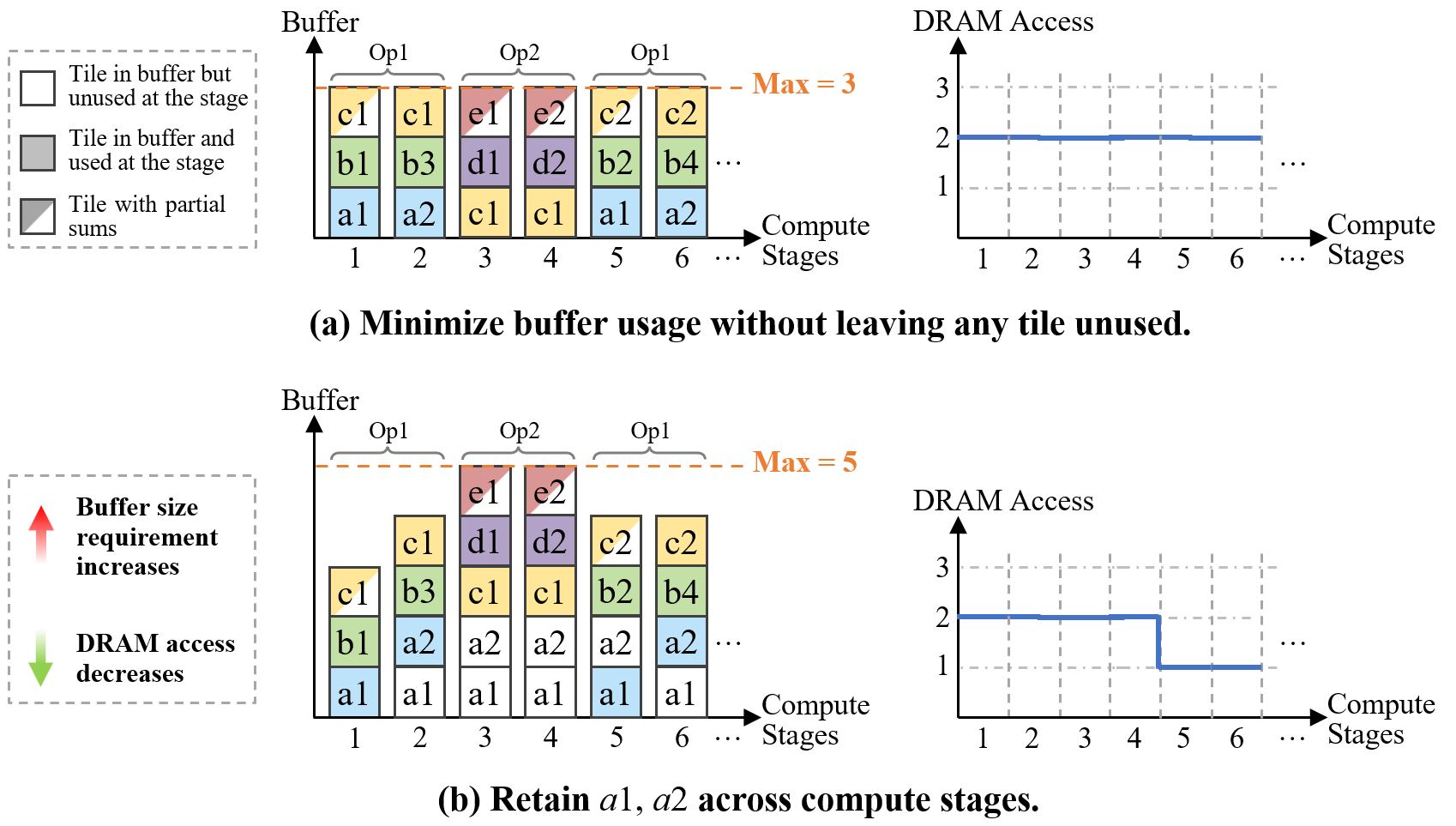}
  \vspace{-2em}
  \caption{Buffer management and its impact. (a) The dataflow follows Fig.~\ref{fig:new_recompute}(a). (b) $a1$ and $a2$ are retained once loaded into the buffer, eliminating repeated DRAM accesses for $a1$ and $a2$ from stage~\textcircled{5} to stage~\textcircled{6}.}
  \label{fig:new_buffer_manage}
\end{figure}

Motivated by this observation, we manage buffer allocation at a fine-grained operand tile level. Specifically, operand tiles may remain in the buffer even when they are not used in certain compute stages, a strategy we refer to as \textbf{buffer retention}. In Fig.~\ref{fig:new_buffer_manage}, dataflow (b) retains two tiles of $A$, $a1$ and $a2$, once they are loaded. As a result, (b) requires more buffer space than (a), but reduces DRAM access.

Crucially, buffer management is tightly coupled with computation ordering. For instance, if we choose to retain $b1$ and $b3$ in Fig.~\ref{fig:new_buffer_manage}(b) without modifying the computation ordering, the buffer requirement still increases to five tiles. However, DRAM traffic from stage~\textcircled{5} to stage~\textcircled{6} does not decrease, as tiles $b2$ and $b4$ still must be loaded.


\section{Pseudo Nested Loop Representation}
\label{sec:nested_loop}

We introduce pseudo nested loops as a compact yet expressive representation for tiled fusion dataflows. This form encapsulates all relevant decision elements and, when paired with our performance model, allows exhaustive enumeration of the decision space. Within this abstraction, loop boundaries, loop order, and buffering levels together define each fusion dataflow in a complete, unique, and concise manner.

\subsection{Pseudo Nested Loop} \label{sec: pseudo nested loop}

\begin{figure}[t]
  \centering
  \includegraphics[width=1.0\linewidth]{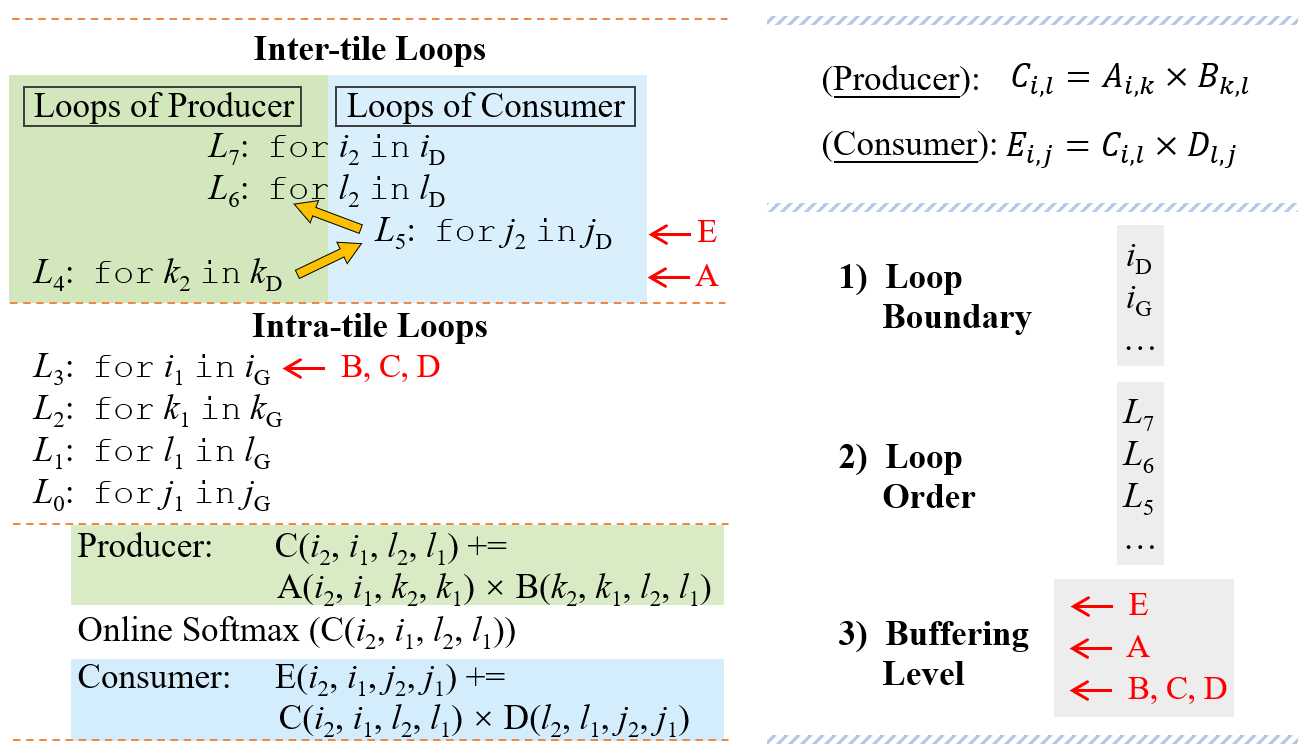}
  \vspace{-2em}
  \caption{Pseudo nested loop example and components. Loop boundaries come from integer factorization (e.g., $I = i_D \times i_G$). Red arrows mark buffering levels; yellow arrows show transition between iteration spaces.}
  \label{fig:new_loop_appearance}
\end{figure}

A generic pseudo nested loop representation is depicted in 
Fig.~\ref{fig:new_loop_appearance}, where the outer four loops correspond to inter-tile iterations. Certain loop layers are shared by both the producer and consumer (e.g., $L_6$ and $L_7$). The inner four loops correspond to intra-tile iterations. Note that the nested structure in Fig.~\ref{fig:new_loop_appearance} is for illustration only and does not reflect the actual implementation, which is not a perfect polyhedron, as discussed in Section~\ref{sec:new_loop_example}.


\textbf{1) Loop Boundary}. 
Tiling is explicitly represented through loop boundaries, such as $i_D$, $i_G$, $l_D$, and $l_G$. Each dimension is tiled based on the integer factorization, e.g., $i_D i_G = I$. These loop boundaries determine the tile sizes of operand matrices and impact the required on-chip buffer capacity.

\textbf{2) Loop Order}. 
Computation ordering is defined by the loop order in the pseudo nested loop. By enforcing the constraint in Section~\ref{sec:new_compute_order}, the computation ordering of any fusion dataflow can be derived precisely from the inter-tile loop sequence, as demonstrated in Section~\ref{sec:new_loop_example}.

\textbf{3) Buffering Level}. 
Buffer management is captured through the buffering level assigned to each operand. Each operand is associated with a specific buffering level, corresponding to one of the loop layers. To represent the buffer-retention behavior described in Section~\ref{sec:new_buffer_manage}, we define the following rule: When the buffering level is assigned within inter-tile loops, a buffer space equal to the operand’s tile footprint is allocated and retained across all applicable computation stages.

\subsection{Learn From an Example} \label{sec:new_loop_example}

\begin{figure*}[t]
  \centering
  \includegraphics[width=1.0\linewidth]{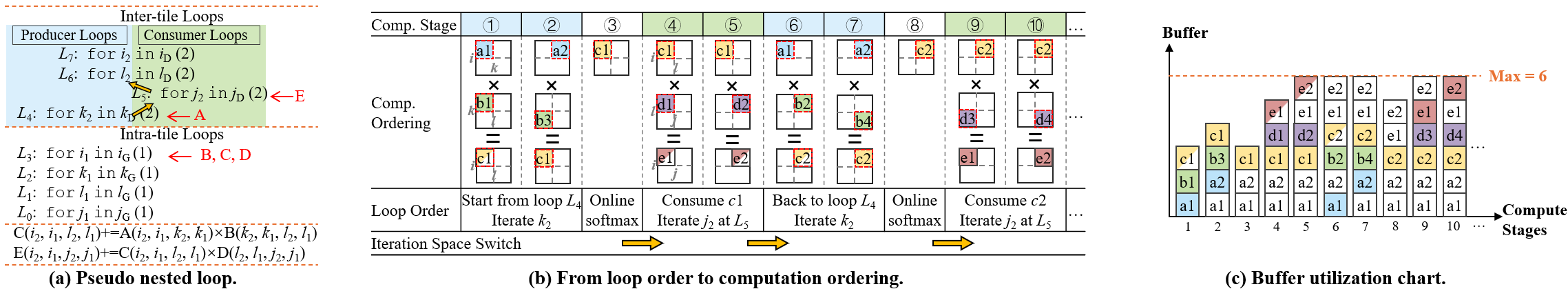}
  \vspace{-2em}
  \caption{(a) Specific numeric values are assigned to loop boundaries to illustrate the \textbf{tiling} decision. (b) The loop is unrolled to reveal the \textbf{computation ordering}. (c) The buffer-utilization chart demonstrates \textbf{buffer management}: buffering levels are assigned to each operand, and the levels for $A$ and $E$ are placed within the inter-tile loops, meaning two tiles of $A$ and two tiles of $E$ are retained in the buffer.}
  \label{fig:new_loop_order_to_compute_order}
\end{figure*}

\textbf{1) Loop Boundary $\rightarrow$ Tiling}. 
In Fig.~\ref{fig:new_loop_order_to_compute_order}(a), numeric values are assigned to all loop boundaries. The minimum data unit for every operand is a $1 \times 1$ tile, and both data movement and computation operate at this tile granularity. Note that a ``tile" here abstracts the actual data volume.

\textbf{2) Loop Order $\rightarrow$ Computation Ordering}. 
In Fig.~\ref{fig:new_loop_order_to_compute_order}(b), unrolling the loop exposes the corresponding computation ordering. In fusion dataflow, the loop structure is not a perfect loop nest (polyhedron); instead, the iteration spaces of different operators may interleave~\cite{zheng2023tileflow}, as highlighted by the yellow arrows in Fig.~\ref{fig:new_loop_order_to_compute_order}(a) and (b).
We begin with loop layer $L_4$, where $k_2$ iterates from 1 to 2, indicating that one tile ($c1$) of the intermediate matrix $C$ is being accumulated. This behavior directly reflects the No Psum Propagation constraint.
After online softmax at stage~\textcircled{3}, $c1$ is consumed from stage~\textcircled{4} to~\textcircled{5}, corresponding to the iteration of $j_2$ at loop layer $L_5$. The iteration space transitions from the producer to the consumer are indicated by the yellow arrows at the bottom of the figure.

Such iteration-space transitions pose a fundamental challenge for representing fusion, distinguishing it from conventional nested loops~\cite{liang2022deep, stoutchinin2019optimally}. The iteration space later switches back to the producer to generate additional intermediate tiles (e.g., $c2$). Once $c2$ is fully accumulated at stage~\textcircled{8}, the consumer processes it from stage~\textcircled{9} to~\textcircled{10}, completing another iteration-space switch.

\textbf{3) Buffering Level $\rightarrow$ Buffer Management}. 
Fig.~\ref{fig:new_loop_order_to_compute_order}(c) shows the resulting buffer behavior from the buffering level assignments in Fig.~\ref{fig:new_loop_order_to_compute_order}(a). Buffering levels for $A$ and $E$ are set inside inter-tile loops, causing buffer retention for their tiles.

For matrix $A$, the buffering level is set at loop layer $L_4$, meaning an entire tile row ($a1$ and $a2$) is retained in the buffer once loaded. Consequently, buffer usage increases from 3 to 4 tiles from stage~\textcircled{1} to stage~\textcircled{2}, holding $a1$, $a2$, $b3$, and $c1$ at stage~\textcircled{2}.
Similarly, the buffering level for matrix $E$ is assigned at loop layer $L_5$, retaining tiles $e1$ and $e2$ once they are loaded. In contrast, matrices $B$, $C$, and $D$ set buffering levels at loop layer $L_3$, discarding one tile once unused. For example, buffer usage decreases from 4 to 3 tiles between stage~\textcircled{2} and stage~\textcircled{3} because $b3$ is discarded after use.
When all compute stages finish, the peak buffer usage across stages directly yields the required on-chip buffer size.

\section{Analytical Performance Model} \label{sec: section V}


In this section, we present a computationally efficient performance model that forms the foundation of MMEE. The model relies on two assumptions: (1) the workload is full dense attention computation; and (2) the dataflow generated is preserved by the compiler backend, as in software-defined hardware approaches such as Groq \cite{abts2022software, 9895630}.

By leveraging pseudo nested loops (Section~\ref{sec:nested_loop}), we significantly extend an intra-operator model~\cite{liang2022deep} to support cross-operator fusion dataflows. The model is entirely analytical, enabling extremely fast estimation of latency and energy, and thereby facilitating very efficient solution search. This stands in contrast to algorithmic models, such as the tree model~\cite{zheng2023tileflow}, which requires multiple steps for each estimation and thus hinder rapid search over a large decision space. 
Although earlier work on DNN performance modeling~\cite{zhao2020dnn} provides analytical formulas, it relies on numerous ``if--else'' branches to select the correct expression for each application scenario. Such branching substantially increases estimation time. In contrast, our model is fully analytical and avoids any ``if--else'' parsing.
For simplicity, we omit the head dimension in all expressions. This does not sacrifice generality, as computations across different heads are independent~\cite{kao2023flat} and can therefore be easily mapped onto separate PE arrays.

\subsection{Terminology} \label{sec:terms}

We define several terms used in our analytical models.

\noindent
\textbf{Operand's dimensions}. 
The operand's dimensions refer to the union of its loop variables, e.g., $\{i_2, k_2, i_1, k_1\}$ for matrix $A$.

\noindent
\textbf{Operator's dimensions}. The operator's dimensions refer to the union of both operands' dimensions for the operator, e.g.,
$\{i_2,\allowbreak k_2,\allowbreak l_2,\allowbreak i_1, k_1, l_1\}$ for the producer operator. 

\begin{figure}[t]
  \centering
  \includegraphics[width=0.90\linewidth]{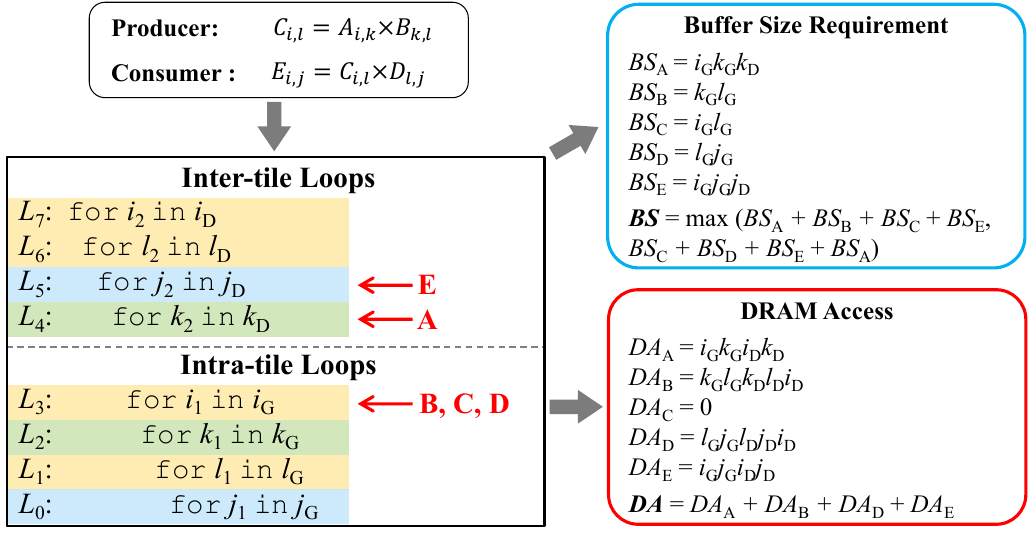}
  \vspace{-1em}
  \caption{Example of buffer-size requirement and DRAM-access estimation. Each expression is a composition of additive and multiplicative combinations of loop boundaries.}
  \label{fig:loop analysis}
\end{figure}

\noindent
\textbf{Operand's effective dimensions}. 
If there is no recomputation or the operator for the operand is a consumer, the effective dimensions of the operand are the same as the operator's dimensions. For the operand of a producer operator in the presence of recomputation, the effective dimensions are the union of the producer operator's dimensions and the corresponding consumer operator's dimensions. For the example in Fig.~\ref{fig:loop analysis}, if there is recomputation, the effective dimensions of $A$ are $\{i_2, k_2, l_2, i_1, k_1, l_1, j_2, j_1\}$.

\noindent
\textbf{Reduction dimension.} For an operator, this is the loop variable shared by its two operands, e.g., $k_2$ for the producer. 

\noindent
\textbf{Blocker}. 
For an operand, its blocker is the innermost loop layer outside its buffering level whose iteration causes the operand’s buffered data to change.
For operand $A$, its buffering level is $L_4$ in Fig.~\ref{fig:loop analysis}.
Its blocker is $L_7$ because it corresponds to one of operand $A$'s dimensions while $L_5$ and $L_6$ have nothing to do with $A$. For an operand of the consumer, its blocker can also correspond to the reduction dimension of its producer. For example, the blocker of operand $D$ is $L_4$.

\subsection{Models for Estimating Buffer Size Requirements} \label{sec:calculate_buf_size}

\noindent
{\bf Operand's buffer size requirement.}
The buffer size required for an operand is determined by the product of the loop boundaries corresponding to its dimensional loops at and below its buffering level. In Fig.~\ref{fig:loop analysis}, the dimensions for operand matrix $A$ based on its buffering level are $k_2$, $i_1$, and $k_1$, whose associated loop boundaries are $k_D$, $i_G$, and $k_G$, respectively. Thus, the buffer size requirement for $A$ is given by $BS_A = k_D  i_G  k_G$.

\noindent
{\bf Operator's buffer size requirement.}
First, we define a binary indicator $\tau_Y$, which is set to 1 if the buffering level of operand $Y$ is at an inter-tile loop, and 0 otherwise. For a given operator $M$, let $\Delta^M$ denote the set of operands exclusively associated with $M$. In the example shown in Fig.~\ref{fig:loop analysis}, we have $\Delta^{\text{Op1}} = \{A, B\}$. The operand shared between the two operators, Op1 and Op2, is denoted as $\Delta^{\text{Op1}, \text{Op2}}$, which in this case is $C$.

For two fused operators, Op1 and Op2, their buffer size requirements can be estimated as follows: 
\begin{align} \label{eq:eq2}
    BS^{Op1} = \sum_{X\in \Delta^{Op1}\cup \Delta^{Op1, Op2}} BS_X + \sum_{Y\in \Delta^{Op2}} \tau_Y BS_Y  \\
     BS^{Op2} = \sum_{X\in \Delta^{Op2}\cup \Delta^{Op1, Op2}} BS_X + \sum_{Y\in \Delta^{Op1}} \tau_Y BS_Y 
\end{align}
Here, $\tau_Y = 1$ indicates that some tiles of operand $Y$ must remain in the buffer even when its associated operator is not currently being executed. For example, in Fig.~\ref{fig:loop analysis}, some tiles of matrix $E$ are buffered during the execution of Op1, even though $E$ is not an operand of Op1. Consequently, the buffer size requirement for Op1 is given by: 
\begin{align} \label{eq:eq3}
    BS^{Op1} & = BS_A + BS_B + BS_C + \tau_D \cdot BS_D + \tau_E \cdot BS_E \notag \\
                  & = BS_A + BS_B + BS_C + BS_E 
\end{align}
This formulation highlights a key difference from the models in \cite{liang2022deep}, which are not designed to account for the complexities introduced by operator fusion. 

\noindent
{\bf Overall buffer size requirement.}
The two fused operators are executed sequentially and therefore the overall buffer size requirement is the maximum between them:
\begin{align} \label{eq:eq5}
    BS = \text{max} (BS^{Op1}, BS^{Op2})
\end{align}
This also differs from the models of \cite{liang2022deep}, where operator fusion is not considered.

\subsection{Models for Estimating DRAM Access} \label{sec:calculate_dram_access}

DRAM access refers to the volume of data transferred between the on-chip buffer and off-chip DRAM. Intermediate results, such as matrix $C$, are not permitted to be written to DRAM, and therefore its DRAM access $DA_C = 0$. The derivation of DRAM access in our approach is significantly more complex than that in \cite{liang2022deep} due to the data dependencies in the fused operators and the presence of recomputation.  

\noindent
{\bf DRAM access of an operand without recomputation:}

\noindent
\underline{Scenario 1:} When the blocker of an operand lies within its own dimensions, the operand's DRAM access is calculated as the product of: (1) its buffer size requirement, (2) the loop boundary of its blocker, and (3) the loop boundaries of all effective dimensions above the blocker. For operand $A$ in Fig.~\ref{fig:loop analysis}, the buffer size requirement is $BS_A = i_G k_G k_D$, the loop boundary of its blocker is $i_D$, and there are no additional dimensions above $i_2$. Thus, the DRAM access for $A$ is estimated as: 
\begin{align} DA_A = BS_A i_D = i_G k_G k_D i_D \end{align} 
This corresponds to the total size of matrix $A$, indicating that each element is fetched from DRAM exactly once.

\noindent
\underline{Scenario 2:} 
When an operand is associated with the consumer operator and its blocker is a reduction dimension of the producer operator, the corresponding DRAM access is estimated as the product of: (1) its buffer size requirement, and (2) the loop boundaries of all effective dimensions above the blocker. Unlike Scenario 1, the loop boundary of the blocker is excluded, as the blocker is not directly related to the operand.
For operand $D$ in Fig.~\ref{fig:loop analysis}, where the blocker is $k_2$, the DRAM access is estimated as: 
\begin{align}
    DA_D = BS_D l_D j_D i_D = l_G j_G l_D j_D i_D
\end{align}
This means the DRAM access equals $i_D$ copies of matrix $D$ (size $l_Gj_Gl_Dj_D$), or each element is accessed $i_D$ times.

\noindent
{\bf DRAM access of an operand with recomputation:}
When producer operands $A$ and $B$ are not retained in the buffer, recomputing the intermediate result $C$ requires reloading the corresponding tiles of $A$ and $B$ from DRAM. This recomputation overhead is already captured in the two scenarios above through effective dimensions. These dimensions change dynamically depending on whether recomputation is applied.

\noindent
{\bf Overall DRAM access:}
\begin{align}
    DA = DA_A + DA_B + DA_D + DA_E
\end{align}

\subsection{Energy Consumption and Latency Estimation} \label{sec:energy_latency_models}

Our energy consumption estimation builds upon existing models~\cite{zheng2023tileflow, parashar2019timeloop, wu2019accelergy}, augmented by our analytical DRAM access models. It includes three components: 1. DRAM - buffer traffic energy, depending on $DA$; 2. buffer - register file traffic energy, influenced by the stationary mode (e.g., weight stationary); and 3. computation energy, proportional to total MAC (Multiply-Accumulate) operations. The total MAC count is given by $N_{op1} + N_{op2}$ without recomputation, or $j_D N_{op1} + N_{op2}$ with recomputation, where $N_{op1}=i k l$ is the MAC count for operator 1 and $N_{op2}=i j l$ for operator 2. We evaluate all combinations of three popular stationary modes (weight, input, and output stationary) for both operators under each computation ordering and buffer management solution.

\begin{sloppypar}
Similarly, latency estimation is based on established frameworks~\cite{huang2024mind, zheng2023tileflow, cai2023inter}, and assumes the data load, execution, and data store are fully pipelined with double buffer. It is estimated as the maximum of DRAM access latency, proportional to $DA$ under a given DRAM bandwidth, and computation latency, determined by the the PE array size and its utilization.
\end{sloppypar}

The energy and latency of softmax are captured in our performance model. We introduce a factor $c_{softmax}$ to account for the energy of softmax operations. This user-defined parameter depends on hardware implementation and accuracy requirements \cite{lin2025systolicattention, wei2020design}. Since only Op1 ($Q \times K^T$) requires softmax, its energy is modeled as $c_{softmax}il$ ($c_{softmax}ilj_D$ for recomputation). For latency, we adopt a tile-level pipeline: the current tile proceeds to SFU immediately after Op1, while the logits matrix ($L$) from the previous tile is multiplied with $V$ to perform $L \times V$. This pipeline enables latency hiding between consecutive tiles, consistent with FlashAttention \cite{shah2024flashattention}. To support this, sufficient SFU resources are allocated \cite{kao2023flat}.

\subsection{Accounting for Loop Order and Buffering Levels} \label{sec:vector_form}

The models described so far are in terms of tile sizes (loop boundaries) for a specific setting of loop order and buffering level. To capture the impact from different loop orders and buffering levels, we adopt the ``$\ln-\exp$" trick presented in \cite{huang2021cosa,liang2022deep}.
Consider a performance metric such as buffer size requirement for operand $A$, $BS_A = k_D i_G k_G$, for the loop order and buffering level shown in Fig.~\ref{fig:loop analysis}. It can be rewritten in vector form as:
\begin{align}
    BS_A=\exp(q\cdot \ln(b)) \label{eq:ln-exp}
\end{align}
where the query vector and boundary vector are defined as:
\begin{align}
    & q = [0, 1, 0, 0, 1, 1, 0, 0] \\
    & b = [i_D, k_D, l_D, j_D, i_G, k_G, l_G, j_G]^T
\end{align}

Likewise, buffer size requirements and DRAM accesses for any loop orders and buffering levels  can be expressed in this form, where the \textbf{boundary vector} encodes the tiling decisions, while the \textbf{query vector} encodes loop ordering and buffering level decisions. Based on analytical formulas like
Equation~\eqref{eq:ln-exp} and the process described in Section~\ref{sec:energy_latency_models}, latency and energy can be estimated for any legitimate loop orders, buffering levels and tiling without the need for ``if-else’’ parsing \cite{zhao2020dnn} or tree traversal \cite{zheng2023tileflow}. Please note that the model of \cite{huang2021cosa,liang2022deep} is restricted to intra-operator dataflows while ours works for cross-operator fusion dataflows due to our significant extensions described in Sections~\ref{sec:calculate_buf_size} and \ref{sec:calculate_dram_access}. 

\section{Dataflow Optimization} \label{sec:dataflow_opt}

\subsection{Optimization by MMEE}

\begin{figure}[t]
  \centering
  \includegraphics[width=0.9\linewidth]{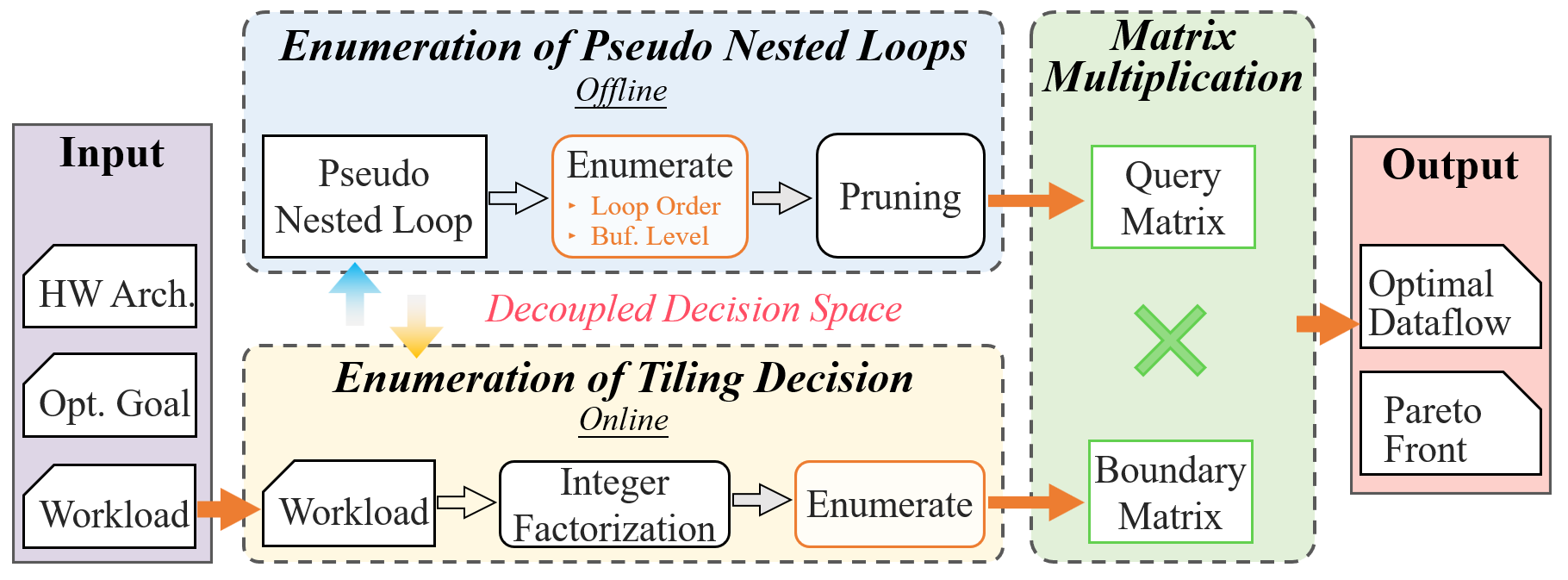}
  \vspace{-1em}
  \caption{Overview of the MMEE dataflow optimization framework. The decision space is decoupled into two independently enumerated subspaces, represented in matrix form. Optimal dataflow solutions and performance tradeoffs are then obtained through matrix multiplication.}
  \label{fig:new_overview}
\end{figure}

An overview of our MMEE (Matrix Multiplication Encoded Enumeration) method is shown in Fig.~\ref{fig:new_overview}. A key idea in MMEE is \underline{partitioning the decision space into two parts}:  
(1) loop orders and buffering levels, whose candidate options are enumerated offline; and  
(2) tiling configurations, whose options are enumerated online.  
The rationale behind offline enumeration is that almost all attention mechanisms share the same pseudo-nested-loop structure, regardless of workload. Consequently, their loop-order and buffering-level options can be enumerated once and reused across many workloads. In contrast, valid tile sizes depend on workload-specific dimensions, making online enumeration necessary for tiling. The number of tiling options is finite because tile sizes must correspond to integer factorizations of the operand matrix dimensions. For example, an operand matrix with $I$ rows can be partitioned into $i_D$ tile rows, each containing $i_G$ rows, such that $I = i_G \cdot i_D$, where both $i_G$ and $i_D$ are positive integers.

This decision space partitioning coincides with the ``$\ln-\exp$"-based model by Equation~\eqref{eq:ln-exp}, where a solution is evaluated by the inner product between the query vector, which encodes loop order and buffering level options, and the boundary vector, which indicates tiling configurations. Then, multiple loop ordering and buffering level solutions can be described by a {\bf query matrix $Q=[q_1, q_2, ..., q_m]^T$}, where each row is a query vector. Likewise, multiple tiling options can be represented by a {\bf boundary matrix $B=[b_1, b_2,...,b_n]$}, where each column corresponds a boundary vector. Then, solutions for all loop order, buffering level and tiling options can be evaluated by the product between $Q$ and $B$ as
\begin{align}
    \text{exp}
    (
    \begin{bmatrix}
    q_1 \\
    q_2 \\
    ... \\
    q_m
    \end{bmatrix}
    \text{ln}
    (
    \begin{bmatrix}
    b_1 & b_2 & ... & b_n
    \end{bmatrix}
    )
    )
    =
    \begin{bmatrix}
    r_{11} & r_{12} & ... & r_{1n} \\
    r_{21} & r_{22} & ... & r_{2n} \\
    ... & ... & ... & ... \\
    r_{m1} & r_{m2} & ... & r_{mn} \\
    \end{bmatrix} \label{eq:MMEE}
\end{align}
where $r_{ij}$ is the result, such as buffer size requirement or DRAM access, of loop order and buffering level corresponding to $q_i$ and tiling configuration $b_j$.


MMEE utilizes six types of query matrices:
$Q_{BS,P}$ for producer buffer size requirements, $Q_{BS,C}$ for consumer buffer size requirements,
$Q_{DA}$ for DRAM access, $Q_{C,P}$ for producer computation latency, 
$Q_{C,C}$ for consumer computation latency,
and $Q_{BR}$ for traffic volume between the buffer and register files. 
In addition, a $Q$ matrix encodes the option of whether recomputation is performed.

By integrating these query matrices with the boundary matrix 
$B$ via Equation~\eqref{eq:MMEE}, and by applying the analytical models and pruning, energy and latency estimates can be computed for all enumerated solutions. This exhaustive enumeration enables the identification of optimal configurations for a variety of objectives. In addition, Pareto fronts can be obtained to show tradeoffs between performance metrics, such as buffer size vs. DRAM traffic (Fig.~\ref{fig:compare with MG}) or energy vs. latency (Fig.~\ref{fig:pareto}).

Although matrix encoding also appears in~\cite{liang2022deep}, there are three key differences.  
(1) ~\cite{liang2022deep} targets only intra-operator dataflows, while MMEE supports cross-operator dataflows. In particular, the six query matrices used in MMEE substantially differ to correctly handle fused operators.  
(2) ~\cite{liang2022deep} relaxes the problem into a continuous domain and requires a discretization step to produce integer solutions, making the optimization inherently heuristic. MMEE instead exhaustively enumerates and guarantees optimality.  
(3) MMEE incorporates a pruning technique, absent in \cite{liang2022deep}, which accelerates enumeration-based search without compromising optimality.

\subsection{Offline Symbolic Pruning} \label{sec:pruning}

Some computation ordering and buffer management solutions can be pruned out without affecting optimality. We propose a symbolic pruning technique such that the pruning is independent of workloads and tiling solutions. Thus, pruning can be performed offline.

We first partition these solutions into groups according to whether or not recomputation is performed (2 options) and a combination of intra-operator stationary options. Each operator has three stationary options (WS, OS and IS; two operators yield 9 combinations. Overall, there are 18 groups. The pruning is carried out within each group individually.

For each group, the pruning is based on two criteria: buffer size and DRAM access. Even without specific tiling solutions, we can derive symbolic expressions for buffer size requirement and DRAM access corresponding to each query matrix row and perform pairwise comparisons. For example, given two buffer management solutions $s_u$ and $s_v$, if their buffer sizes are estimated as $BS_{u}=i_Gk_G$ and $BS_{v}=i_Gk_Gi_D$, then $BS_u < BS_v$ holds for all valid tiling solutions. For any solution pair $(s_u, s_v)$, $s_v$ is inferior to $s_u$ and pruned if  
\begin{align} \label{eq:eq12}
  & BS_v \ge BS_u   ~~\text{and}~~ DA_v > DA_u, \nonumber \\
 \text{or}~~ & BS_v > BS_u   ~~\text{and}~~ DA_v \ge DA_u 
\end{align}

Applying the pruning reduces each query matrix for each group from 20K rows to 58, revealing the large redundancy in the computation-ordering and buffer-management subspace.

\subsection{Optimality of MMEE} \label{sec:optimality proof}

\noindent
{\bf Statement:} MMEE guarantees to find the optimal latency-energy tradeoff for fused attention dataflow mapping according to the models described in Section~\ref{sec: section V}.

\begin{proof}
We will show that if a solution $s_v$ is inferior according inequalities~\eqref{eq:eq12} then it must lead to an inferior solution in terms of energy and latency.

First, consider the energy part. Two solutions in the same group cost the same PE computation energy (as both have the same choice of recomputation) and the same buffer-register file traffic energy as they share the same intra-operator stationary option. Then, energy difference between them is determined by the DRAM access energy and DRAM-buffer energy. Since $s_v$ must be worse in both DRAM access and buffer size, $s_v$ must be inferior in the overall energy, considering that DRAM-buffer energy is proportional to buffer size.

Second, consider latency, which is the maximum between between computation latency and DRAM access latency. Solutions in the same group have identical computation latency. So the difference is decided by DRAM latency, i.e., DRAM access volume divided by bandwidth. Since $s_v$ has higher DRAM access, then it must have inferior latency in the same group.

Overall, a solution pruned according to inequalities~\eqref{eq:eq12} is inferior in terms of both energy and latency in its group. 
Since all solutions are enumerated and only the energy-latency inferior solutions are pruned out, the pruning would retain all optimal solutions in terms of energy-latency tradeoff.

\end{proof}

\section{Evaluation} \label{sec:experiments}



\subsection{Experiment Setup} \label{sec:experimental setup}

To evaluate MMEE across different hardware settings, we use two accelerator configurations. Accel.~1 follows the NVDLA design \cite{nvdla2017, zheng2023tileflow}, featuring 4 PE arrays, a 1 MB on-chip buffer, 60 GB/s DRAM bandwidth, and $32 \times 32$ PEs per array. Accel.~2 adopts a TPU-like design \cite{seshadri2022evaluation, jouppi2017datacenter}, with 4 PE arrays, a 4 MB buffer, 128 GB/s DRAM bandwidth, and $128 \times 128$ PEs per array. Both accelerators operate at 1 GHz.
Softmax follows FlashAttention \cite{milakov2018online, dao2022flashattention, dao2023flashattention}, with $c_{softmax} = 10$.
Energy parameters (SRAM access and PE compute) are taken from \cite{yang2020interstellar} under 28 nm. All experiments are conducted on an AMD Ryzen 7 7840H CPU @ 3.8GHz.

\subsection{Model Validation} \label{sec:model validation}

\begin{figure}[t]
\centering
    \begin{minipage}[b]{0.49\columnwidth}
      \centering
      \includegraphics[width=0.9\linewidth]{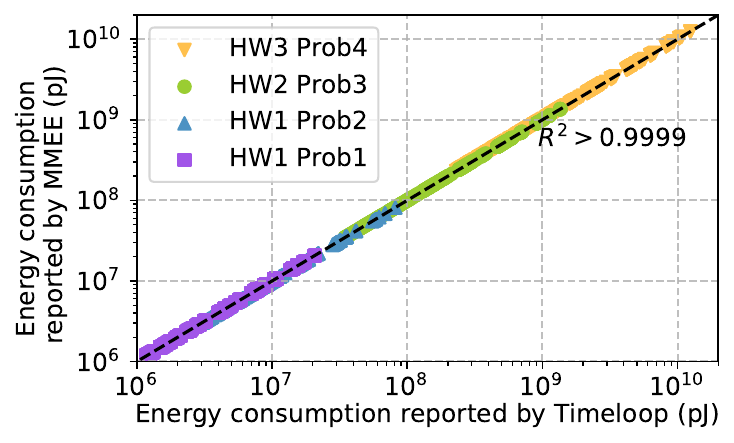}
      \parbox{\linewidth}{\centering \scriptsize \textbf{(a) Energy consumption.}}
    \end{minipage}
\hfill
    \begin{minipage}[b]{0.49\columnwidth}
      \centering
      \includegraphics[width=0.9\linewidth]{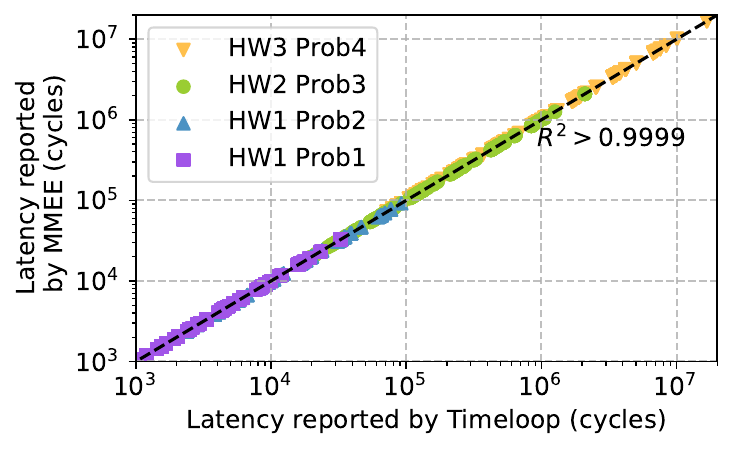}
      \parbox{\linewidth}{\centering \scriptsize \textbf{(b) Latency.}}
    \end{minipage}
\vspace{-2em}
\caption{Our model vs. Timeloop.}
\label{fig:timeloop}
\end{figure}

We validate our models for latency, energy, buffer size, and DRAM access using Timeloop \cite{parashar2019timeloop}, a widely adopted reference. A total of 1410 diverse intra-operator mappings are evaluated across three hardware configurations (HW1–HW3) and four matrix multiplication workloads (Prob1–Prob4), covering a broad performance range. Fig.~\ref{fig:timeloop} shows the results.
Our models match Timeloop closely, achieving $R^2 > 0.9999$ for both latency and energy. Mean errors are near zero, with {\bf the maximum errors} of only 0.5\% in energy and 0.05\% in latency. For reference, TileFlow~\cite{zheng2023tileflow} also validated its model against Timeloop and reported similarly high accuracy.

\begin{figure}[t]
\centering
    \begin{minipage}[b]{0.49\columnwidth}
      \centering
      \includegraphics[width=0.9\linewidth]{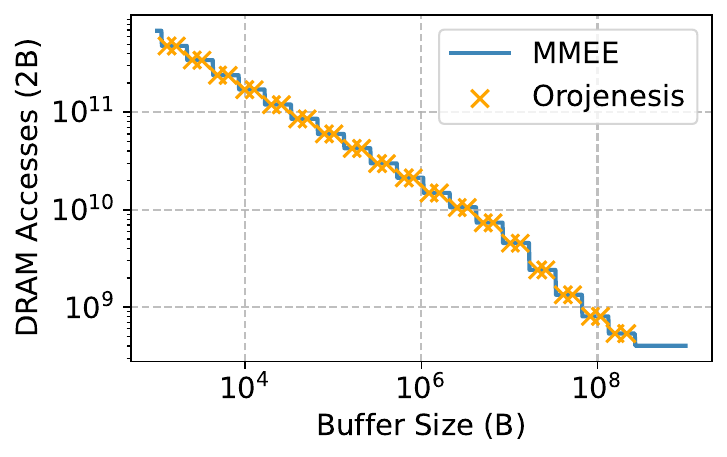}
      \parbox{\linewidth}{\centering \scriptsize \textbf{(a) Fusing 32k\_4k\_16k and \\ 32k\_16k\_4k GEMMs.}}
    \end{minipage}
\hfill
    \begin{minipage}[b]{0.49\columnwidth}
      \centering
      \includegraphics[width=0.9\linewidth]{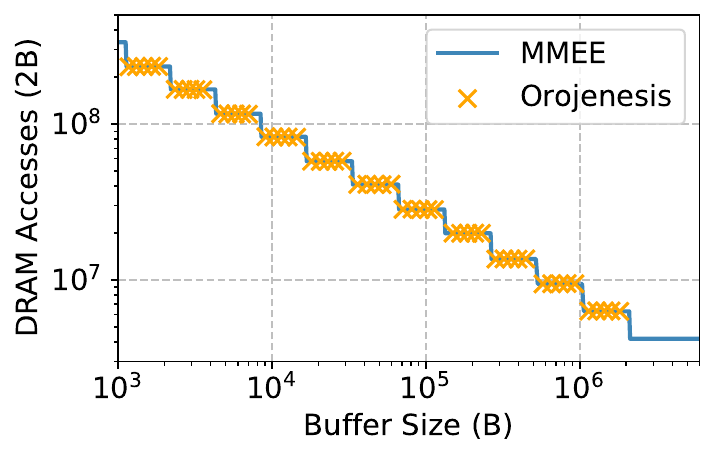}
      \parbox{\linewidth}{\centering \scriptsize \textbf{(b) Fusing 1k\_1k\_1k and \\ 1k\_1k\_1k GEMMs.}}
    \end{minipage}
\vspace{-1.5em}
\caption{DRAM access and buffer size requirement estimation comparison between our model and Nvidia's Orojenesis.}
\label{fig:orojenesis}
\end{figure}

We also compare our models with
NVIDIA's Orojenesis~\cite{huang2024mind}, which estimates DRAM access and buffer size requirement for fusion dataflow mapping.
Fig.~\ref{fig:orojenesis} shows the comparison results for two workloads. 
The results show that mean errors are 0.33\% and 0.25\%, while {\bf maximum errors} are 0.78\% and 0.68\% for the two workloads, respectively.

\subsection{Dataflow Optimization for DRAM Access} \label{sec:compare with orojenesis}

\noindent
\textbf{Baselines}. Two baselines are used: Orojenesis~\cite{huang2024mind} and a no-fusion approach. The no-fusion baseline applies intra-operator optimization independently. To analyze MMEE’s sources of improvement, we also include two variants: ``O+BM" (Orojenesis enhanced with buffer management) and ``O+BM+Re" (Orojenesis further enhanced with recomputation).

\noindent
\textbf{Workloads}. Following Orojenesis, we evaluate fused feedforward networks (FFN) of GPT-3-6.7B. In addition, since attention benefits most from fusion \cite{kao2023flat}, we compare Orojenesis, MMEE, and the two variants on fused attention in GPT-3-6.7B.

\begin{figure}[t]
  \centering
  \includegraphics[width=\linewidth]{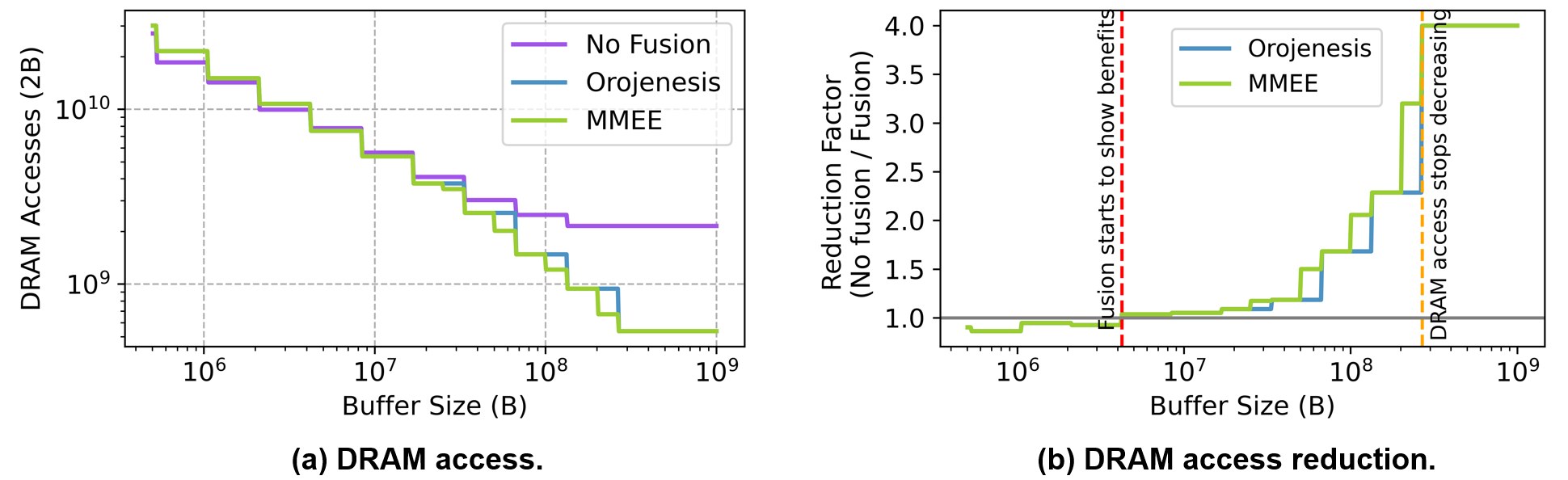}
  \vspace{-2em}
\caption{Fusing FFN networks of GPT-3-6.7B.}
\label{fig:compare with MG}
\end{figure}

\begin{figure}[t]
  \centering
  \includegraphics[width=\linewidth]{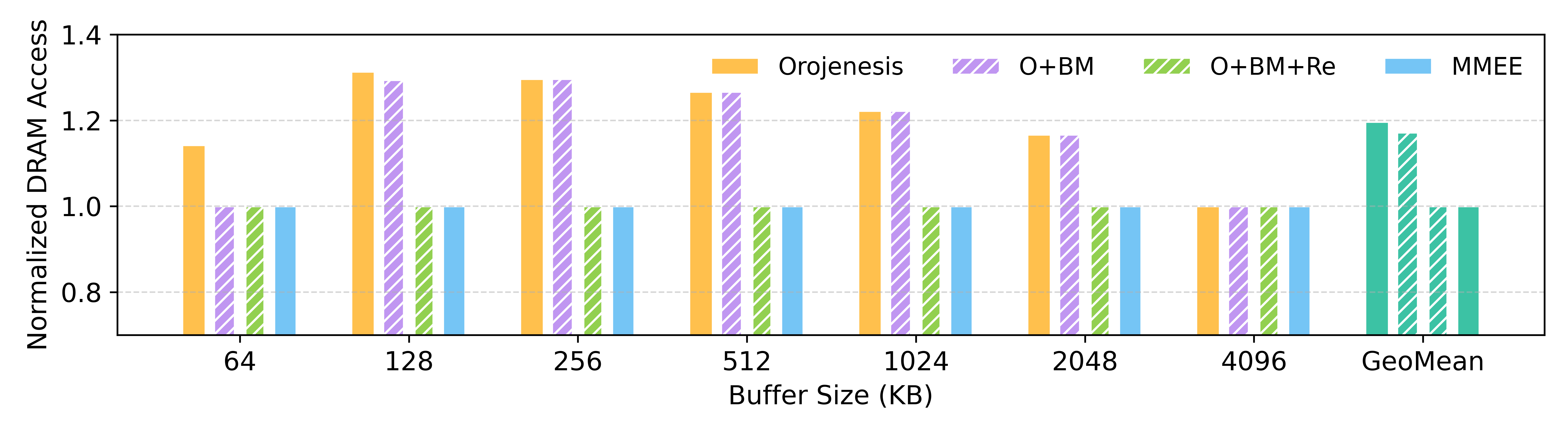}
  \vspace{-2.5em}
\caption{Fusing attention computation of GPT-3-6.7B.}
\label{fig:compare with MG 2}
\end{figure}

\noindent
\textbf{DRAM access vs. buffer size requirement}. 

\noindent
\underline{\textbf{Fusing FFN.}} Fig.~\ref{fig:compare with MG} compares DRAM access among MMEE, Orojenesis, and the no-fusion baseline across various buffer size requirements. The red and orange lines in Fig.~\ref{fig:compare with MG}(b) highlight a region where fusion outperforms no-fusion. On average, our MMEE achieves a $1.5 \times$ DRAM access reduction relative to the no-fusion baseline, surpassing the $1.39 \times$ improvement of Orojenesis by an additional factor of $1.08\times$. 
At specific design points, the gap widens. For example, MMEE achieves a $1.27\times$ reduction at 30 MB (typical in cloud accelerators \cite{jouppi2017datacenter, aws_neuroncore_v2, kao2023flat}) and $1.30\times$ at 1 MB (designs in \cite{cai2023inter, seshadri2022evaluation}).

\noindent
\underline{\textbf{Fusing Attention.}} Fig.~\ref{fig:compare with MG 2} shows DRAM access of MMEE across buffer sizes from 64 KB to 4 MB \cite{nvdla2017, zheng2020efficient, gong2025crane}, achieving up to $1.30 \times$ reduction. ``O+BM" and ``O+BM+Re" provide sources of improvement. Buffer management reduces DRAM traffic by 1.14$\times$ at 64 KB, while recomputation achieves 1.20-1.31$\times$ reduction at larger buffer sizes. At 4 MB, the buffer can hold all attention matrices, yielding no difference among the four mappers.

Overall, MMEE’s performance gains stem from: (1) a larger exploration space that includes buffer management and recomputation; and (2) exhaustive search.

\noindent
\textbf{Runtime}. For the fusion task in Fig. \ref{fig:compare with MG}, Orojenesis evaluates $7.2 \times 10^6$ mappings in around 1200 seconds.
In contrast, MMEE evaluates $7.6 \times 10^8$ mappings within 3.5 seconds, making it \textbf{343× faster} than Orojenesis. This speedup comes from: (1) decoupling the decision space and precomputing all combinations of computation ordering and buffer management, and recomputation; (2) matrix-based parallel evaluation; (3) pruning that further reduces evaluation cost.

\begin{figure*}[t]
\centering
    \begin{minipage}[b]{0.49\linewidth}
      \centering
      \includegraphics[width=\linewidth]{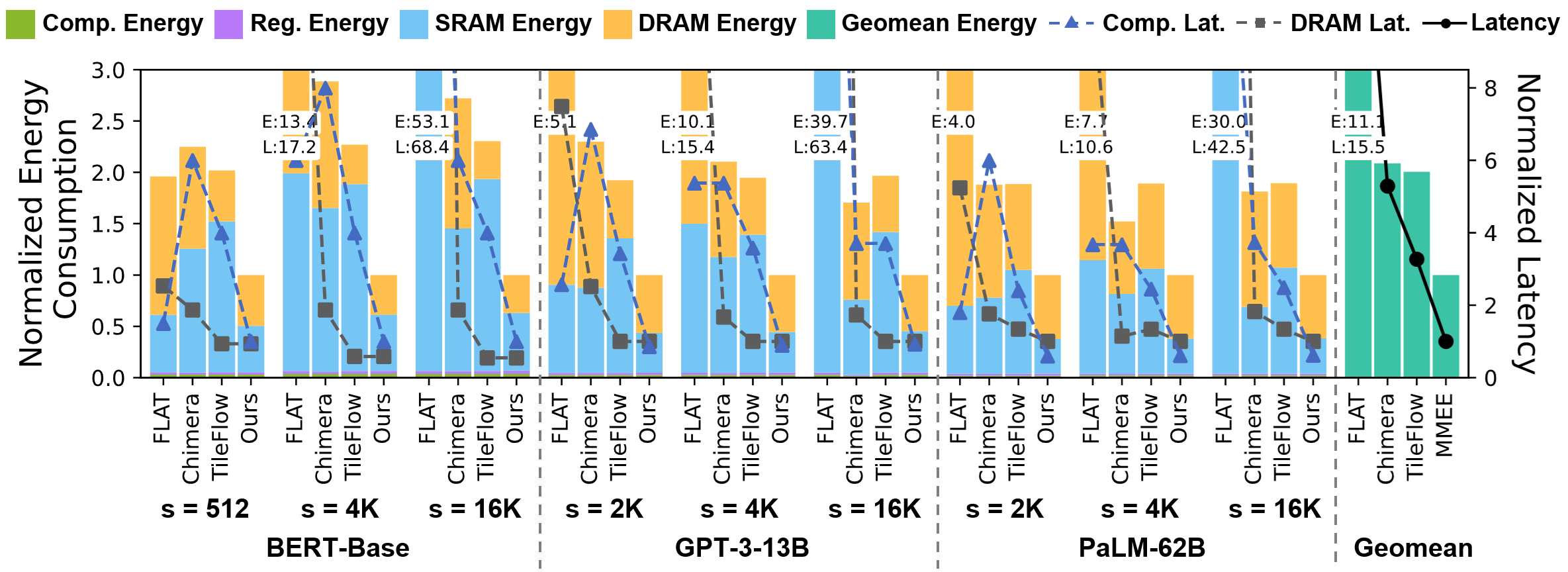}
      \parbox{\linewidth}{\centering \scriptsize \textbf{(a) Energy-driven optimization (E-Driven).}}
    \end{minipage}
\hfill
    \begin{minipage}[b]{0.49\linewidth}
      \centering
      \includegraphics[width=\linewidth]{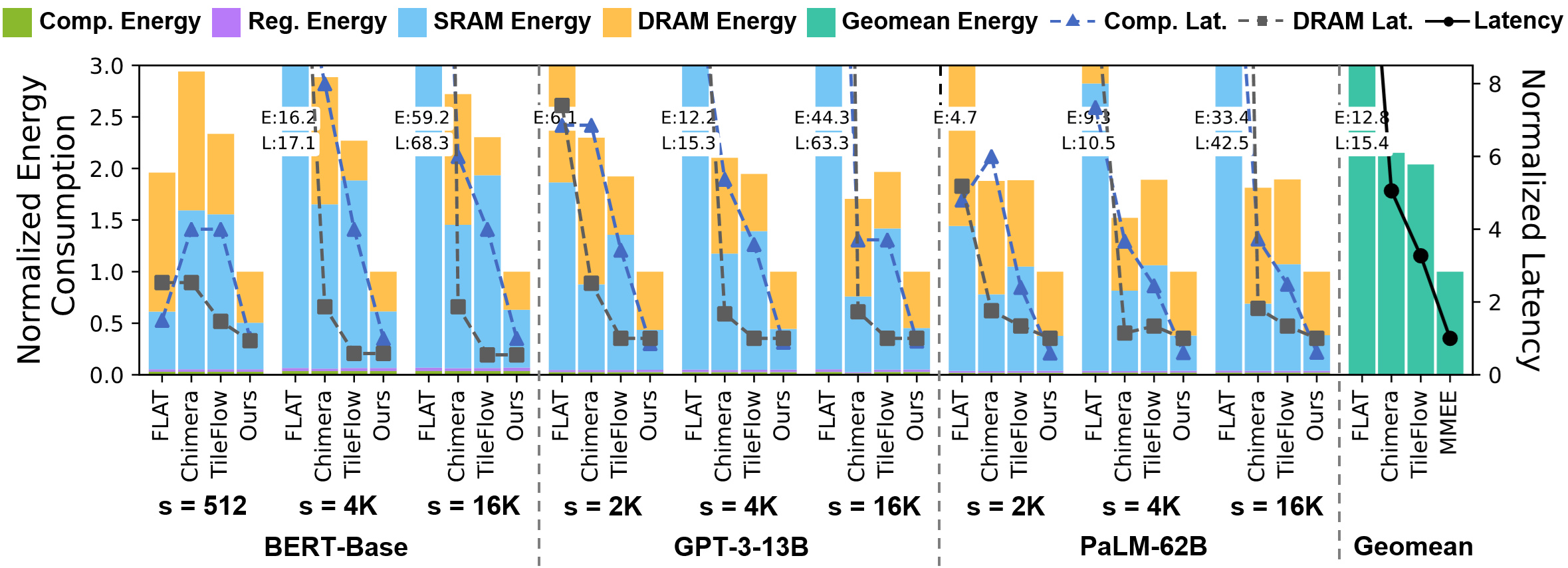}
      \parbox{\linewidth}{\centering \scriptsize \textbf{(b) Latency-driven optimization (L-Driven).}}
    \end{minipage}
\vspace{-0.8em}
\caption{Dataflow optimization results on energy and latency for Accel. 1. The overall energy is the sum of four components, and the overall latency is determined by the larger value between computation latency (Comp. Lat.) and DRAM latency (DRAM Lat.).}
\label{fig:compare with Tileflow on edge}
\end{figure*}

\begin{table}[t]
\centering
\caption{Absolute energy consumption and latency values for MMEE in Figs. \ref{fig:compare with Tileflow on edge} and \ref{fig:compare with Tileflow on cloud}. Each mapping solution is represented in energy/latency (\textnormal{mJ/ms}). Results for other methods can be derived using the relative ratios in the figures.}
\setlength{\tabcolsep}{5pt}
\scriptsize
\label{tab:energy and latency values}
\begin{tabular}{cccc|cc}
\noalign{\hrule height 1pt}
\multicolumn{1}{l}{\multirow{2}{*}{\textbf{Model}}} & \multicolumn{1}{l}{\multirow{2}{*}{\textbf{\begin{tabular}[c]{@{}l@{}}Seq\\ Len\end{tabular}}}} & \multicolumn{2}{c|}{\textbf{Accel. 1 (mJ/ms)}} & \multicolumn{2}{c}{\textbf{Accel. 2 (mJ/ms)}} \\ \cline{3-6} 
\multicolumn{1}{l}{} & \multicolumn{1}{l}{} & E-Driven & L-Driven & E-Driven & L-Driven \\ \noalign{\hrule height 1pt}
\multirow{3}{*}{\begin{tabular}[c]{@{}c@{}}BERT\\ -Base\end{tabular}} & 512 & 1.11/0.10 & 1.11/0.10 & 0.92/0.03 & 0.92/0.03 \\ \cline{2-6} 
 & 4K & 57.14/6.29 & 57.14/6.29 & 38.34/0.79 & 45.33/0.54 \\ \cline{2-6} 
 & 16K & 890.60/100.66 & 890.60/100.66 & 588.49/12.58 & 700.33/6.88 \\ \hline
\multirow{3}{*}{\begin{tabular}[c]{@{}c@{}}GPT\\ -3-13B\end{tabular}} & 2K & 129.75/12.23 & 129.75/12.23 & 65.50/1.80 & 65.50/1.80 \\ \cline{2-6} 
 & 4K & 505.86/46.84 & 505.86/46.84 & 248.14/6.23 & 248.14/6.23 \\ \cline{2-6} 
 & 16K & 7936.0/724.2 & 7936.0/724.2 & 3804.0/87.8 & 3804.0/87.8 \\ \hline
\multirow{3}{*}{\begin{tabular}[c]{@{}c@{}}PaLM\\ -62B\end{tabular}} & 2K & 268.68/27.96 & 268.68/27.96 & 132.03/4.98 & 180.58/3.93 \\ \cline{2-6} 
 & 4K & 1060.8/109.6 & 1060.8/109.6 & 505.9/18.3 & 700.1/14.2 \\ \cline{2-6} 
 & 16K & 16807/1727 & 16807/1727 & 7829/275 & 10936/208 \\ \noalign{\hrule height 1pt}
\end{tabular}
\end{table}

\subsection{Dataflow Results on Energy and Latency} \label{sec:dataflow solution quality}


\noindent
\textbf{Baselines}. We compare our approach against three state-of-the-art cross-operator dataflow mappers FLAT~\cite{kao2023flat}, Chimera~\cite{zheng2023chimera} and TileFlow~\cite{zheng2023tileflow}. Specifically, we select R-Gran from FLAT~\cite{kao2023flat}, which offers the most fine-grained dataflow suited for limited on-chip buffer resources. Chimera~\cite{zheng2023chimera} is not open-sourced and is reproduced using TileFlow's implementation~\cite{zheng2023tileflow}. Orojenesis is excluded as it does not report energy or latency results.

For a fair comparison with TileFlow, we clone its official GitHub implementation and evaluate both TileFlow and MMEE on Accel.~1 and Accel.~2 under identical hardware settings (Section~\ref{sec:experimental setup}). TileFlow must be configured as either energy-driven or latency-driven before search, whereas MMEE evaluates all dataflows and metrics simultaneously. For consistency, we group MMEE results into the same two modes.
TileFlow’s reported runtime only measures MCTS-based tiling search, and buffer management and computation ordering decisions are pre-searched by a genetic algorithm and fixed in the released code. We use TileFlow’s default timeout configuration, which guarantees convergence and matches the settings reported in its paper.

\begin{figure*}[h]
    \centering
    \begin{minipage}{0.49\linewidth} 
        \centering
        \includegraphics[width=\linewidth]{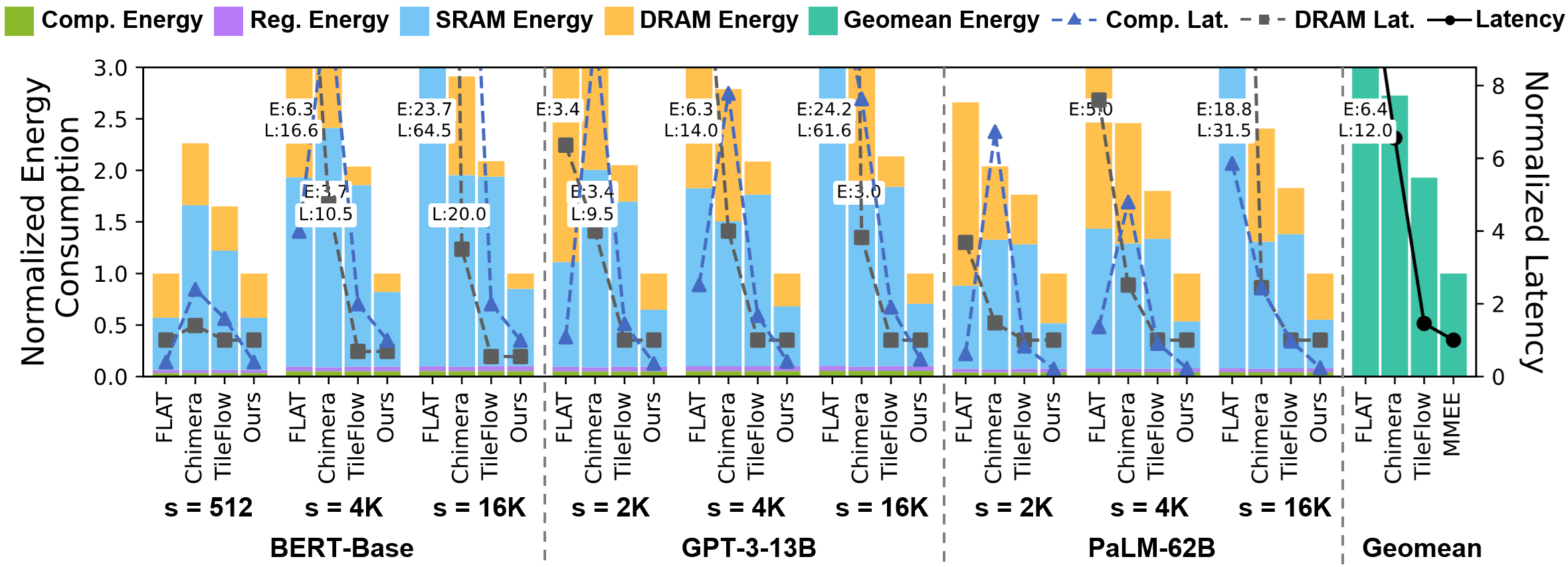}
        \parbox{\linewidth}{\centering \scriptsize \textbf{(a) Energy-driven optimization (E-Driven).}}
    \end{minipage}
    \hfill
    \begin{minipage}{0.49\linewidth}
        \centering
        \includegraphics[width=\linewidth]{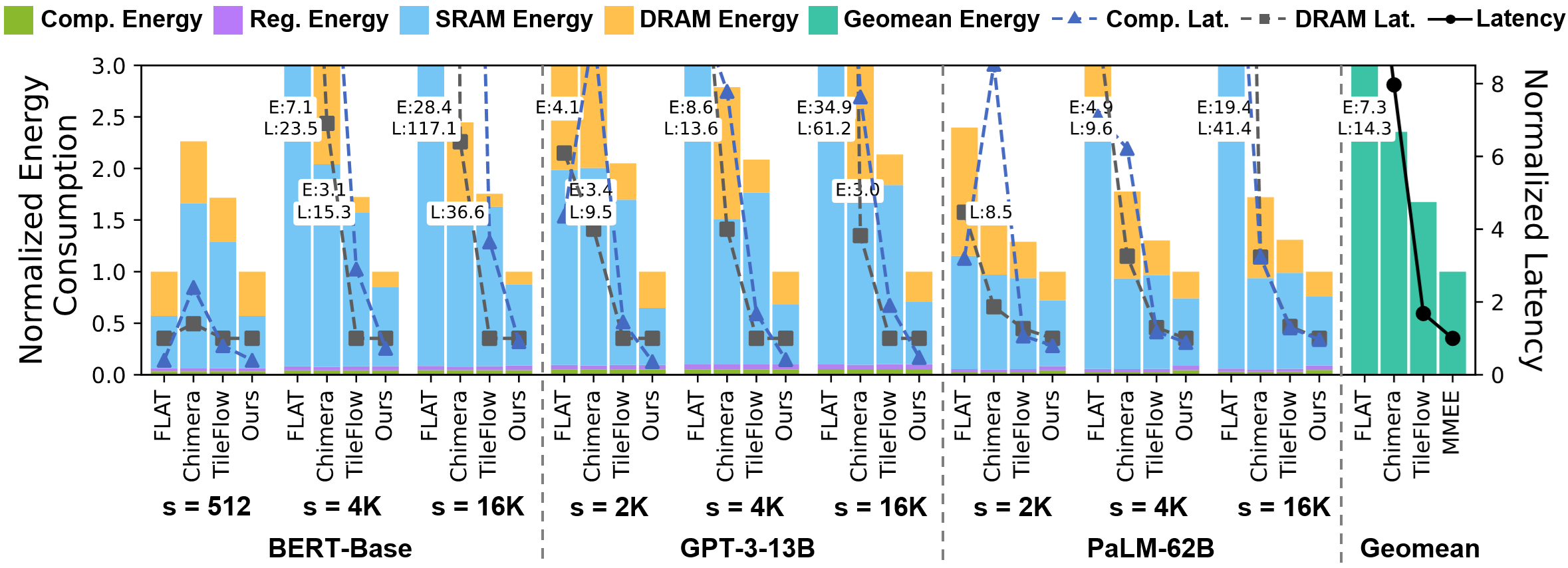}
        \parbox{\linewidth}{\centering \scriptsize \textbf{(b) Latency-driven optimization (L-Driven).}}
    \end{minipage}
    \caption{Dataflow optimization results on energy and latency for Accel. 2.}
    \vspace{-1.2em}
    \label{fig:compare with Tileflow on cloud}
\end{figure*}

\noindent
\textbf{Workloads.} We evaluate MMEE on attention workloads in training/prefill, which use matrix-form queries and exhibit quadratic complexity. Unlike typically memory-bound decoding workloads \cite{kwon2023efficient, ho2024block} that focus on KV-cache access and memory reduction \cite{kwon2023efficient, liu2024kivi, sheng2023flexgen, zhang2023h2o}, training/prefill has higher arithmetic intensity and requires effective dataflow mapping to mitigate both compute- and memory-bound behaviors (Figs.~\ref{fig:compare with Tileflow on edge} and \ref{fig:compare with Tileflow on cloud}).

Attention computations from three models of varying scales are used to evaluate the effectiveness of the proposed methods. From small to large, the selected models are BERT-Base \cite{devlin2019bert}, GPT-3-13B \cite{brown2020language} and PaLM-62B \cite{chowdhery2023palm}. In addition to each model's standard sequence length, longer sequences are also explored to assess the scalability of the evaluated methods.

\noindent
\textbf{Energy and latency.} Figs.~\ref{fig:compare with Tileflow on edge} and ~\ref{fig:compare with Tileflow on cloud} present results for energy consumption and latency on Accel. 1 and Accel. 2. To clarify the sources of energy improvement and differentiate between compute-bound and memory-bound scenarios, the corresponding energy and latency breakdowns are provided. Absolute values are reported in Table \ref{tab:energy and latency values}.

\noindent
\underline{Accel. 1.} When optimizing for either energy consumption or latency, MMEE achieves average reductions of 50\% in energy consumption and 69\% in latency, compared to TileFlow.

\noindent
\underline{Accel. 2.} In energy-driven mode, MMEE achieves an average 48\% energy reduction and an average 31\% latency reduction, relative to TileFlow. When optimizing latency, MMEE shows 40\% less energy and 40\% less latency on average.

The energy breakdowns in Figs. \ref{fig:compare with Tileflow on edge} and \ref{fig:compare with Tileflow on cloud} highlight the sources of  MMEE’s energy savings. Among the dominant components, SRAM and DRAM energy, MMEE consistently achieves equal or lower consumption compared to other methods. 
For example, while TileFlow occasionally matches MMEE’s DRAM energy (e.g., BERT-Base-512 on Accel. 1), it incurs significantly higher SRAM energy due to its reliance on a fixed and less efficient template, derived from genetic algorithms, for managing data movement between SRAM and register files. Moreover, for PaLM-62B-16K, TileFlow consumes more DRAM energy, stemming from its limited buffer management exploration and lack of recomputation.

\begin{figure}[t]
  \centering
  \includegraphics[width=0.9\linewidth]{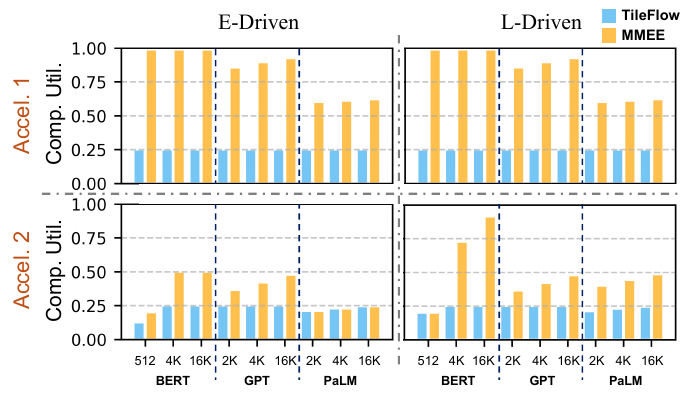}
  \vspace{-1em}
  \caption{Compute utilization of TileFlow and MMEE.}
  \label{fig:compute utilization}
\end{figure}

The latency improvements of MMEE stem from two key factors: higher compute utilization and reduced DRAM access. Compute utilization \cite{kao2023flat} measures how closely a dataflow approaches its arithmetic optimum.  As shown in Fig. \ref{fig:compute utilization}, MMEE achieves substantially higher compute utilization than TileFlow. TileFlow’s low utilization arises from its MCTS-based tiling search, which under-utilizes PE arrays, particularly on Accel.~1 (Fig.~\ref{fig:compare with Tileflow on edge}). This under-utilization leads to only 25\% compute utilization (Fig.~\ref{fig:compute utilization}) and longer latency in compute-bound cases. In memory-bound scenarios, e.g., PaLM-62B models under latency-driven optimization on Accel. 2, MMEE reduces DRAM access due to recomputation, which directly translates to lower overall latency.

\noindent
\textbf{Runtime.} We compare the runtime of TileFlow and MMEE across all experiments shown in Figs.~\ref{fig:compare with Tileflow on edge} and ~\ref{fig:compare with Tileflow on cloud}. TileFlow is configured separately for latency-driven and energy-driven modes, and the reported runtime aggregates results from three models. Runtime results for Chimera and FLAT are not available, as neither their code is open-sourced nor is their runtime reported.
Overall, our approach achieves significant runtime improvements over TileFlow, providing speedups of \textbf{64×} and \textbf{287×} on Accel. 1 and Accel. 2, respectively.




\subsection{Energy-Latency Trade-off} \label{sec:tradeoff between energy and latency}



\begin{figure}[t]
    \centering
    \begin{minipage}{0.49\columnwidth} 
        \centering
        \includegraphics[width=\linewidth]{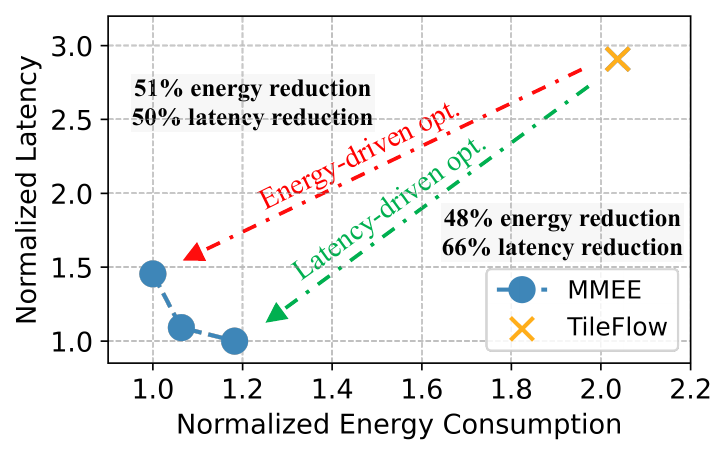}
        \parbox{\linewidth}{\centering \scriptsize \textbf{(a) BERT-Base-4096.}}
    \end{minipage}
    \hfill
    \begin{minipage}{0.49\columnwidth}
        \centering
        \includegraphics[width=\linewidth]{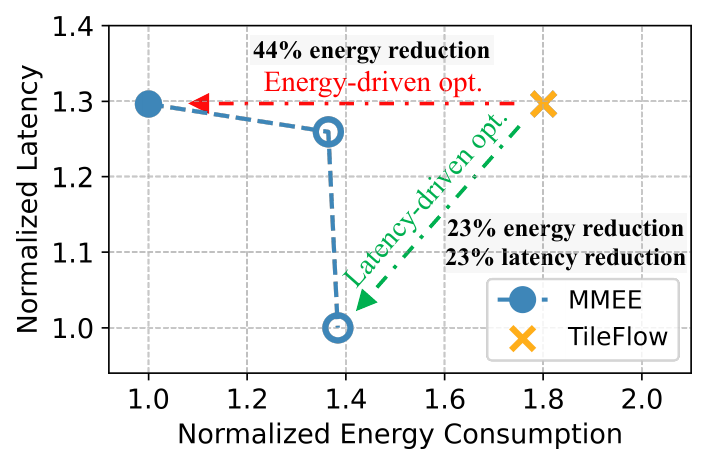}
        \parbox{\linewidth}{\centering \scriptsize \textbf{(b) PaLM-62B-4096.}}
    \end{minipage}
    \vspace{-1em}
    \caption{Energy-latency trade-off of the attention fusion on Accel. 2.}
    \label{fig:pareto}
\end{figure}

\noindent
\textbf{Baseline and workload}. TileFlow is configured with energy-driven and latency-driven optimizations, providing baseline comparisons for evaluating energy-latency trade-offs. We evaluate the fusion of attention operations in BERT-Base and PaLM-62B models, each with a sequence length of 4096.

\noindent
\textbf{Trade-off results}. Fig.~\ref{fig:pareto} shows the Pareto front (blue dashed lines) obtained by MMEE. Red and green arrows denote energy- and latency-driven solutions, respectively. Blue filled and unfilled circles represent recomputation-disabled and recomputation-enabled solutions, respectively. We have three key observations:
(1) Although approximately $10^9$ dataflows are evaluated, \underline{the Pareto front remains sparse}.
(2) Points along the Pareto curve are \underline{highly sensitive}, meaning optimizing energy solely increases latency, and vice versa.
(3) \underline{Recomputation reduces latency} and expands the Pareto frontier, though its impact varies by models. In our experiments, recomputation accounts for two-thirds of the Pareto-optimal points for PaLM models across different sequence lengths, but provides little benefit for BERT-Base and GPT models.

\subsection{Evaluation on GPU} \label{sec:gpu results}


To assess the efficiency of dataflow mappers on real hardware, we deploy the dataflows generated by MMEE and TileFlow on an NVIDIA A100-40GB GPU, using Triton to compile them into CUDA kernels~\cite{tillet2019triton}. The official Triton implementation is used to reproduce FlashAttention-2 (FA2)~\cite{dao2023flashattention, tillet2020triton_tutorials}. 
We adopt FA2 as the primary baseline since it represents the state-of-the-art implementation for the Ampere architecture and is the default attention backend in vLLM~\cite{kwon2023efficient} and SGLang~\cite{zheng2024sglang}. Although FlashAttention-3 (FA3)~\cite{shah2024flashattention} was recently introduced, its improvements primarily target Hopper-specific hardware features and do not introduce algorithmic dataflow advantages over FA2.
For fair comparison, auto-tuning during compilation is firstly disabled. Table~\ref{tab:gpu results} summarizes the results. Overall, MMEE achieves a $2.56\times$ speedup over TileFlow and a $1.18\times$ speedup over FA2, due to its more flexible buffer management and more efficient tiling strategies. 

When auto-tuning is enabled (Auto), higher performance may be achieved by exploiting hardware-specific execution features such as warp scheduling and register allocation \cite{tillet2019triton, li2025tritonbench, kao2023flat}. These factors are managed by the GPU runtime and are not exposed at the dataflow level.
In contrast, MMEE optimizes compile-time controllable architectural decisions, such as tiling and computation ordering, and buffer management, which are primary performance determinants in accelerator architectures \cite{aws_neuroncore_v2, parashar2019timeloop, li2025tl}. When such architectural decisions are fully explored, MMEE consistently achieves lower latency compared to FA2 and TileFlow.

\begin{table}[t]
\centering
\caption{Performance (\textnormal{ms}) of MMEE on an A100 GPU.}
\setlength{\tabcolsep}{5pt}
\scriptsize
\label{tab:gpu results}
\begin{tabular}{cccccccccc}
\noalign{\hrule height 1pt}
Model & \multicolumn{3}{c}{BERT-Base} & \multicolumn{3}{c}{GPT-3-13B} & \multicolumn{3}{c}{PaLM-62B} \\ \cline{2-10} 
Seq Len & 512 & 4K & 16K & 2K & 4K & 16K & 2K & 4K & 16K \\ \hline
TileFlow & 0.05 & 0.64 & 2.45 & 0.55 & 1.14 & 4.24 & 1.63 & 3.23 & 12.83 \\
FA2 & 0.03 & 0.19 & 0.75 & 0.40 & 0.73 & 2.43 & OOM & OOM & OOM \\
Auto & 0.03 & 0.07 & 0.72 & 0.11 & 0.20 & 0.78 & 0.45 & 0.90 & 3.54 \\
MMEE & 0.03 & 0.18 & 0.69 & 0.27 & 0.60 & 2.05 & 0.45 & 0.89 & 3.54 \\ \hline
\end{tabular}
\end{table}

\subsection{Analysis of Energy-Latency Improvement} \label{sec:attribute study}

To analyze the sources of MMEE’s improvements, we decompose the three requirements in Fig.~\ref{fig:previous_work} into two factors: decision space coverage and search efficiency enabled by a branch-free performance model. Since TileFlow uses heuristic search, we equip it with enumeration (TF+) to match MMEE’s search power, isolating whether its limitations arise from its decision space or search algorithm.

\begin{figure}[t]
  \centering
  \includegraphics[width=\linewidth]{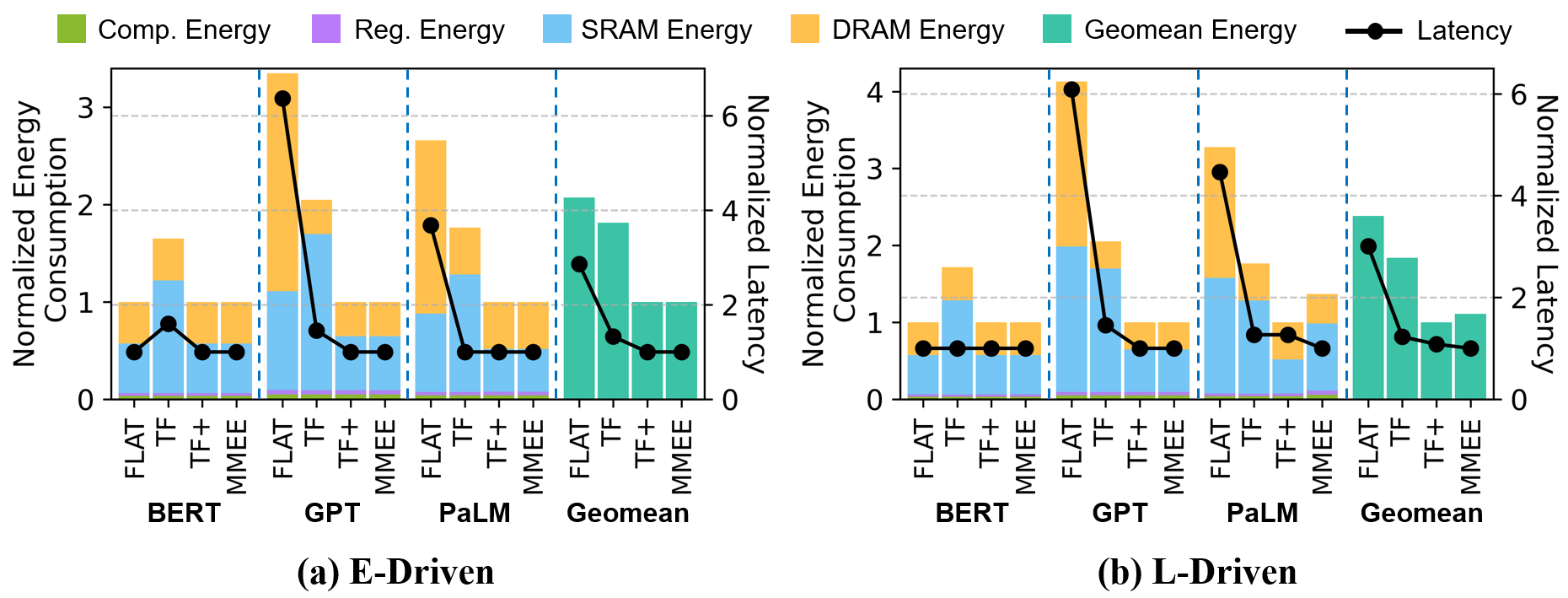}
  \vspace{-2em}
  \caption{Fused attention of various models at base sequence lengths on Accel.~2: (a) energy-driven; (b) latency-driven.}
  \label{fig:attribute study}
\end{figure}

Figs.~\ref{fig:attribute study}(a) and (b) show energy and latency improvements under energy- and latency-driven objectives, respectively. For energy-driven optimization, TF+ matches MMEE’s, whereas TF does not.
This confirms that TileFlow is limited by inefficient heuristic search. Despite using enumeration, FLAT’s limited decision space degrades both energy and latency in several cases (e.g., GPT). Under latency-driven optimization, MMEE’s recomputation enables more flexible tradeoffs, achieving lower latency at slightly higher energy than TF+. Overall, both decision space and search efficiency are critical, with impact varying by workload, objective, and hardware configuration.

\subsection{Runtime Scalability} \label{sec:runtime analysis}

\begin{figure}[t]
  \centering
  \includegraphics[width=\linewidth]{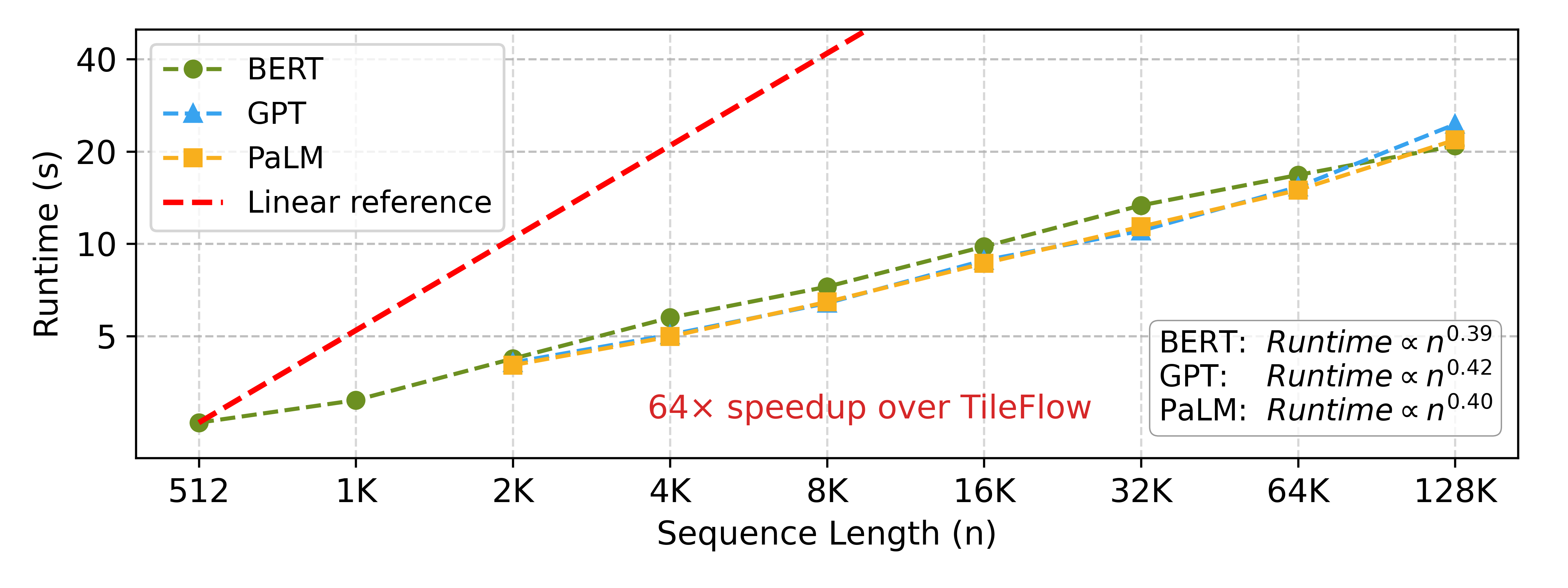}
  \vspace{-2.5em}
  \caption{Runtime of MMEE across varying sequence lengths (log-log).}
  \label{fig:runtime comparison}
\end{figure}

Fig.~\ref{fig:runtime comparison} shows MMEE’s runtime and scaling with sequence length on Accel.~1. MMEE scales sub-linearly; at 128K, runtime remains below 25 seconds. Power-law fitting further quantifies scalability. Due to decision-space decoupling and an efficient performance model, online runtime is dominated by tiling enumeration (integer factorization). For attention workloads, factorization complexity ($\propto n^{0.4}$ on average) grows much slower than quadratic compute complexity.


\subsection{Sensitivity Analysis} \label{sec:sensitivity}

\subsubsection{Sensitivity to sequence lengths} \label{sec:sensitivity to sequence lengths}

\begin{figure}[t]
  \centering
  \includegraphics[width=\linewidth]{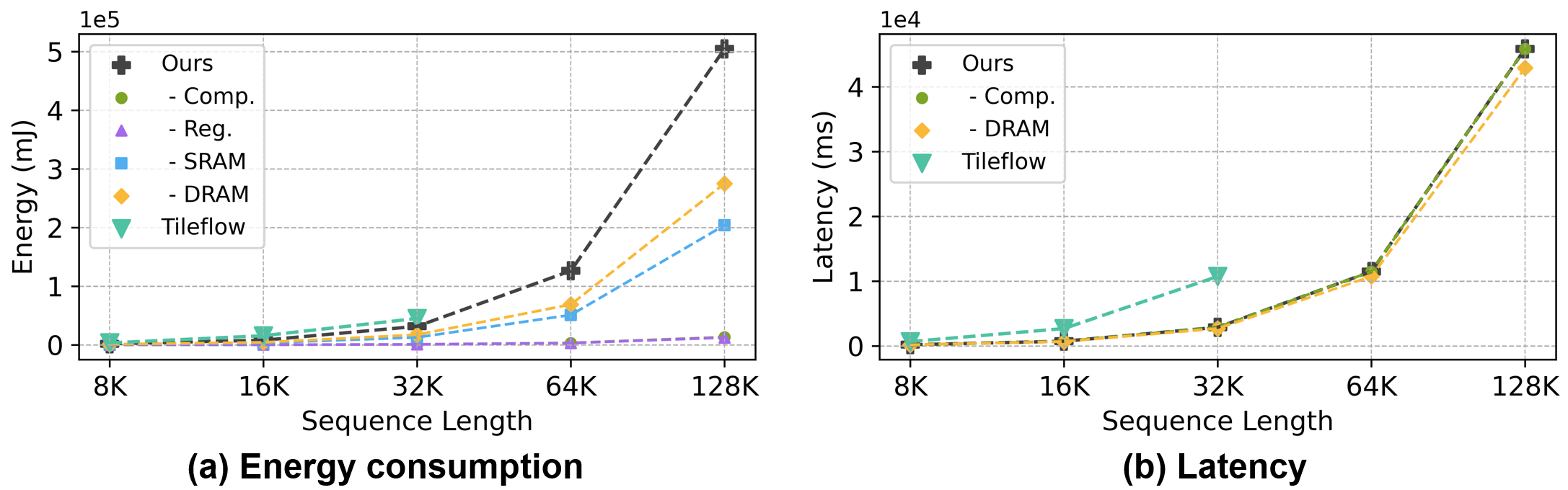}
  \vspace{-2em}
  \caption{Energy and latency trends with increasing sequence length for fused GPT-3-13B attention under energy-driven optimization on Accel. 1.}
  \label{fig:sensitivity to seq len}
\end{figure}

As sequence lengths scale beyond 16K \cite{stallone2024scaling, feng2024long}, we evaluate MMEE’s effectiveness and energy–latency trends under optimal dataflows. Fig.~\ref{fig:sensitivity to seq len} compares MMEE’s energy and latency breakdown with TileFlow’s overall metrics from 8K to 128K. TileFlow is limited to 32K due to code issues. From 8K to 32K, MMEE consistently achieves the lowest energy and latency.
Both energy and latency grow nearly quadratically with sequence length, consistent with attention complexity \cite{keles2023computational} and TileFlow results \cite{zheng2023tileflow}. In the energy breakdown, SRAM and DRAM dominate; recomputation and softmax are included in compute energy. Under the given hardware and workloads, longer sequences do not change the compute-bound nature but increase pressure on compute units and DRAM bandwidth.

\subsubsection{Sensitivity to key decision elements} \label{sec:sensitivity to decision elements}
\begin{figure}[t]
  \centering
  \includegraphics[width=\linewidth]{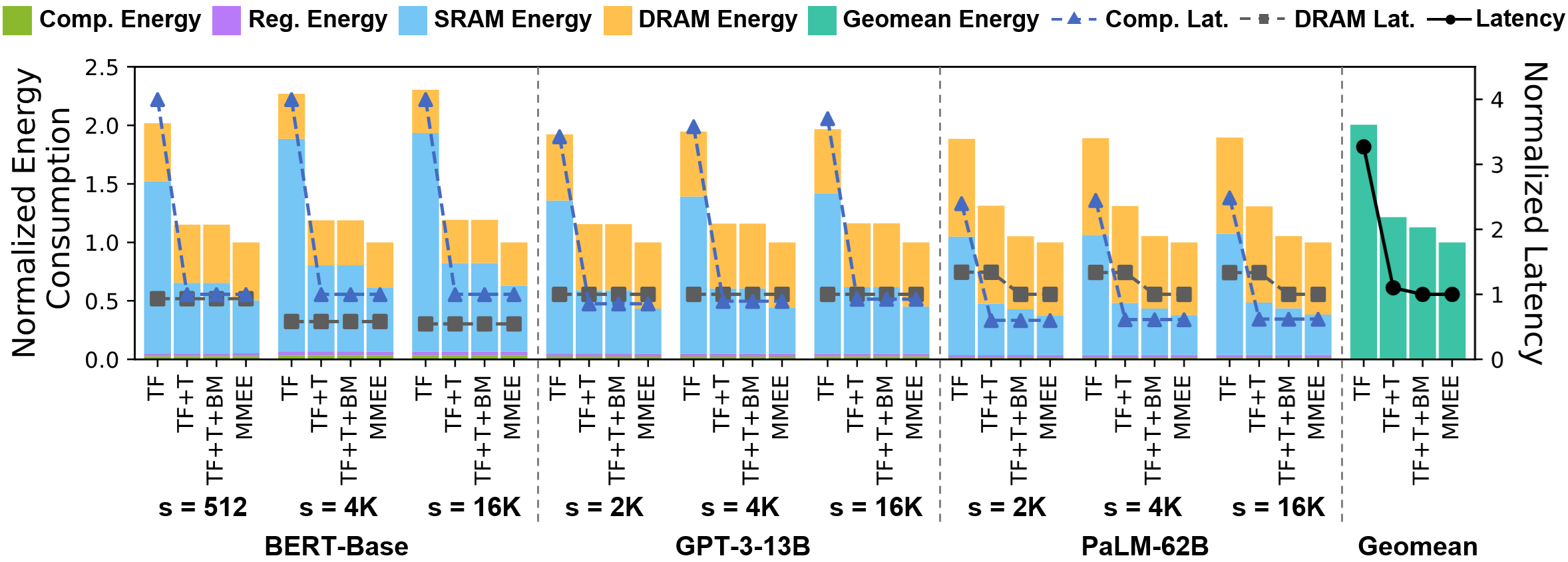}
  \vspace{-2em}
  \caption{Analysis of decision elements. ``TF", ``TF+T", and ``TF+T+BM" denote TileFlow, TileFlow enhanced with tiling enumeration, and TileFlow enhanced with both tiling and buffer management enumeration, respectively. Fusion is under the energy-driven optimization on Accel. 1.}
  \label{fig:sensitivity to decision elements}
\end{figure}

Fig.~\ref{fig:sensitivity to decision elements} quantifies the impact of three key decision elements on dataflow quality. We implement two TileFlow variants within MMEE framework: ``TF+T" and ``TF+T+BM". As discussed in Section~\ref{sec:dataflow solution quality}, TileFlow primarily suffers from an inefficient tiling search. Consequently, ``TF+T" achieves 39\% lower energy and 66\% lower latency than TileFlow. Building upon this, flexible buffer management further reduces energy and latency by 7\% and 9\%, respectively. Enabling computation-ordering exploration yields an additional 11\% energy reduction.

\begin{figure}[t]
  \centering
  \includegraphics[width=\linewidth]{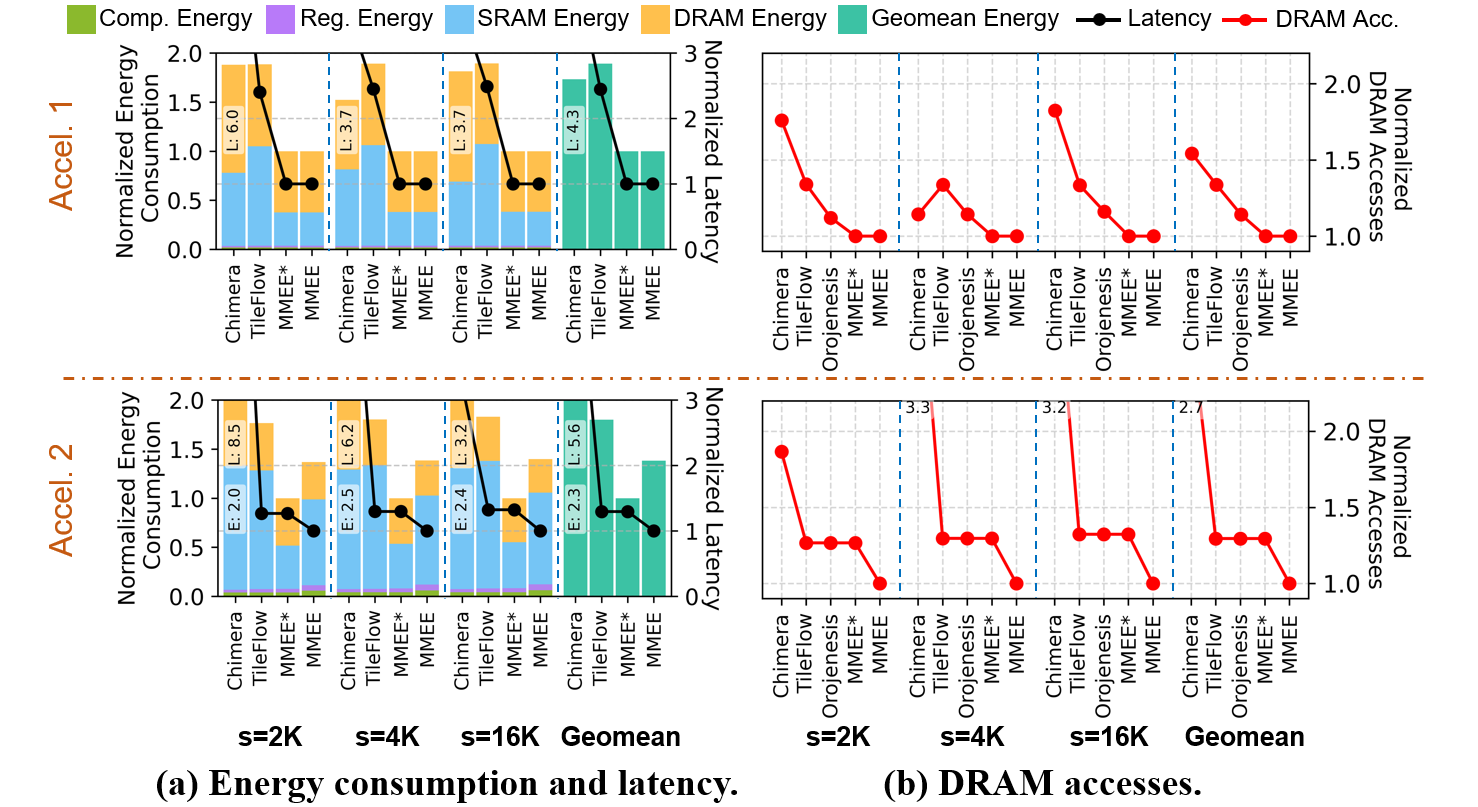}
  \vspace{-2em}
  \caption{Comparison of Chimera, TileFlow, Orojenesis, MMEE* (MMEE without recomputation), and MMEE in terms of energy, latency, and DRAM access. The attention of PaLM-62B is fused under latency-driven optimization.}
  \label{fig:sensitivity to recompute}
\end{figure}

\begin{figure}[t]
  \centering
  \includegraphics[width=\linewidth]{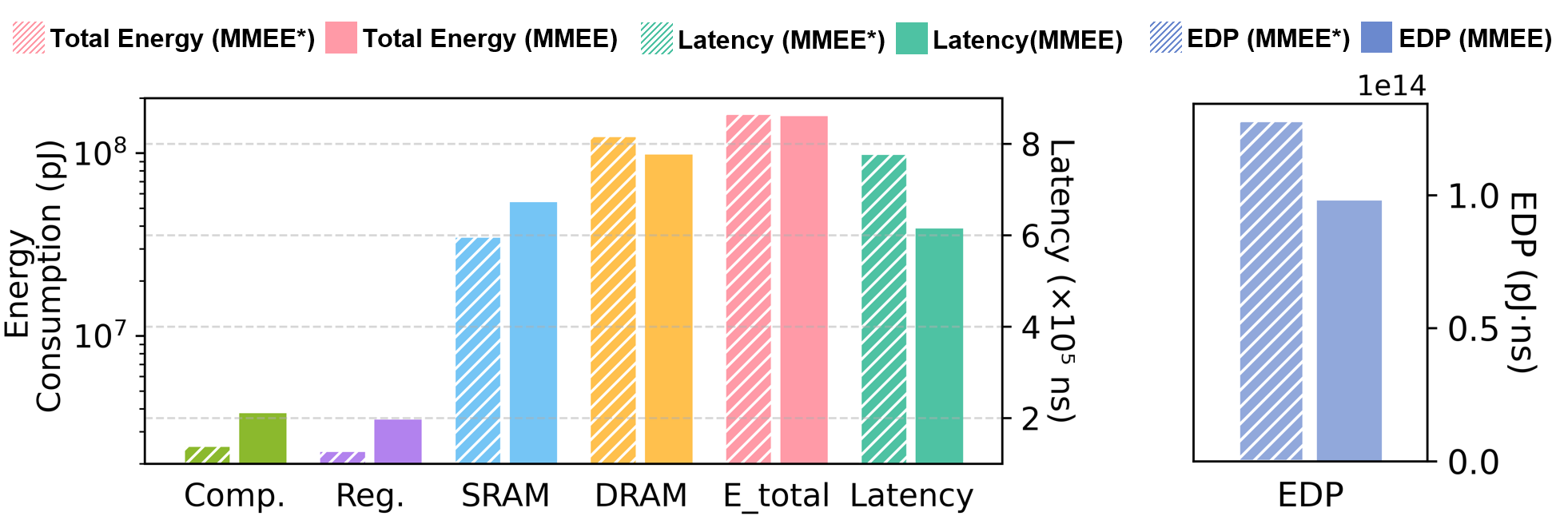}
  \vspace{-2em}
  \caption{A case study of MMEE* (MMEE without recomputation) and MMEE in terms of energy, latency, and EDP.}
  \label{fig:recompute case study}
\end{figure}

\subsubsection{Sensitivity to recomputation} \label{sec:sensitivity to recomputation}
Fig. \ref{fig:sensitivity to recompute} compares MMEE* (MMEE without recomputation) and MMEE across various sequence lengths. Orojenesis is mapping-independent and only reports minimum DRAM access.
On Accel. 1, MMEE* consistently outperforms all baselines in energy, latency, and DRAM access. This shows that the gains are from exhaustive exploration of tiling and buffer management. For example, MMEE* reduces DRAM access by 1.14× compared to Orojenesis.

For Accel. 2, recomputation provides noticeable latency reductions across all experiments. Although MMEE incurs higher energy consumption than MMEE*, it still outperforms other mappers. The increase in energy arises from additional computation and register-to-SRAM data movement. However, the reduced DRAM access leads to improved latency in these memory-bound scenarios. Compared to MMEE*, MMEE reduces both latency and DRAM access by $1.30\times$.

Fig.~\ref{fig:recompute case study} shows a dataflow benefiting from recomputation. We fuse BERT-Base attention with standard sequence length on an industrial accelerator \cite{coral_intro_guide}. Energy is plotted in log scale and latency in linear scale. Compared to no recomputation, MMEE enables recomputation, increasing compute, register file, and SRAM energy. However, it reduces DRAM access, lowering DRAM energy and keeping total energy nearly unchanged. As the workload is memory-bound, reduced DRAM access directly lowers latency, yielding a $1.31 \times$ reduction in EDP.

\begin{sloppypar}
\subsubsection{Sensitivity to pruning} \label{sec:sensitivity to pruning}
In Section~\ref{sec:pruning}, pruning accelerates dataflow exploration. To verify its effectiveness, we repeat all experiments in Figs.~\ref{fig:compare with Tileflow on edge} and \ref{fig:compare with Tileflow on cloud} without pruning. The results show that pruning does not compromise optimality, while achieving significant speedups of $347\times$ on Accel. 1 and $221\times$ on Accel. 2, compared to MMEE without pruning.
\end{sloppypar}

\subsubsection{Sensitivity to hardware design} \label{sec:sensitivity to hardware designs}
To demonstrate the generality of MMEE, we evaluate it on three hardware designs from both industry and academia \cite{coral_intro_guide, zheng2020efficient, cai2023inter, gong2025crane}, as listed in Table~\ref{tab:various hardware designs}. Entries for TileFlow and MMEE report normalized energy and latency values. Compared to TileFlow, MMEE achieves substantial reductions in energy and latency.

\begin{table}[t]
\centering
\setlength{\tabcolsep}{5pt}
\scriptsize
\caption{Performance of MMEE across different hardware designs.}
\label{tab:various hardware designs}
\begin{tabular}{c|cccccc}
\noalign{\hrule height 1pt}
HW Design & \begin{tabular}[c]{@{}c@{}}PE Array\\ Size\end{tabular} & \begin{tabular}[c]{@{}c@{}}\#PE \\ Array\end{tabular} & \begin{tabular}[c]{@{}c@{}}Buffer\\ Size\end{tabular} & \begin{tabular}[c]{@{}c@{}}DRAM\\ Bandwidth\end{tabular} & \begin{tabular}[c]{@{}c@{}}TileFlow\\ (E/L)\end{tabular} & \begin{tabular}[c]{@{}c@{}}MMEE\\ (E/L)\end{tabular} \\ \noalign{\hrule height 1pt}
Coral \cite{coral_intro_guide}  & 16$\times$16 & 1 & 32 KB & 1.6 GB/s & 1.95/1.59 & 1/1 \\
Design \cite{zheng2020efficient} & 32$\times$32 & 1 & 512 KB & 2 GB/s & 2.24/1.18 & 1/1 \\
SET \cite{cai2023inter, gong2025crane} & 32$\times$32 & 16 & 16 MB & 8 GB/s & 4.17/2.56 & 1/1 \\ \noalign{\hrule height 1pt}
\end{tabular}
\end{table}

\subsection{Applicability to Fusion Patterns Beyond Attention} \label{sec:applications of MMEE}

MMEE can optimize dataflows for workloads with two convolutions or two GEMMs. Table~\ref{tab:applications of MMEE} reports results on Accel.~1. For convolution chains, convolutions are converted to GEMMs via im2col, with shapes defined by input height/width, input channels, output channels of both convolutions, and kernel sizes. For two GEMMs, shapes are $[I, K, L, J]$. The baseline is the better result between TileFlow and intra-operator optimization. Overall, MMEE consistently achieves the best energy–latency tradeoffs.

\begin{table}[t]
\centering
\setlength{\tabcolsep}{5pt}
\scriptsize
\caption{MMEE Performance for Convolution Chains and Two GEMMs}
\label{tab:applications of MMEE}
\begin{tabular}{ccccc}
\noalign{\hrule height 1pt}
\multicolumn{2}{c}{Workload} & Problem Shape & Baseline (E/L) & MMEE (E/L) \\ \hline
\multirow{2}{*}{\begin{tabular}[c]{@{}c@{}}Conv.\\ Chain\end{tabular}} & CC1 \cite{zheng2023tileflow} & {[}$112^2$,64,192,128,$3^2$,$1^2${]} & 2.34/1.16 & 1/1 \\
 & CC2 \cite{zheng2023tileflow} & {[}$56^2$,64,64,64,$1^2$,$1^2${]} & 1.20/1.50 & 1/1 \\ \noalign{\hrule height 1pt}
\multirow{2}{*}{\begin{tabular}[c]{@{}c@{}}Two \\ GEMMs\end{tabular}} & MLP \cite{zheng2023chimera} & {[}768,64,384,64{]} & 1.93/1.00 & 1/1 \\
 & FFN \cite{devlin2019bert} & {[}2048,768,3072,768{]} & 1.08/1.14 & 1/1 \\ \noalign{\hrule height 1pt}
\end{tabular}
\end{table}

\begin{figure}[t]
  \centering
  \includegraphics[width=\linewidth]{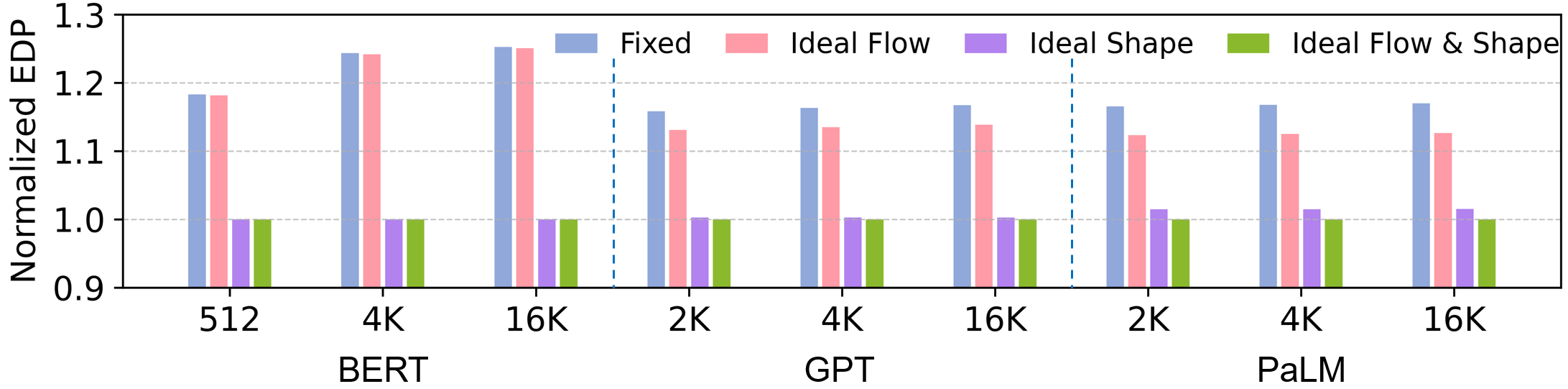}
  \vspace{-2em}
  \caption{EDP-driven optimization of reconfigurable PE arrays. Fixed: 32×32 WS array. Ideal Flow: reconfigurable stationary modes on a 32×32 array. Ideal Shape: flexible PE shaping under WS with $\leq$ 32×32 total PEs. Ideal Shape \& Dataflow: joint reconfiguration of PE shape and stationary mode.}
  \label{fig:dynamic hardware application}
\end{figure}

\subsection{Applicability to Reconfigurable PE Arrays} \label{sec:dynamic hardware application}

MMEE supports exploration of reconfigurable arrays. We consider two reconfiguration dimensions: stationary mode and logical array shape \cite{lee2024resa, han2024redas, wang2022novel, wang2025desa}. The baseline uses weight-stationary on a 32 $\times$ 32 array; other settings follow Accel.~1. Multiple array shapes are evaluated by MMEE. Fig.~\ref{fig:dynamic hardware application} compares four array designs under EDP-driven optimization. Array reshaping provides greater benefits than stationary-mode flexibility, though optimal shapes vary by workloads.

\subsection{Discussion} \label{sec:scope of MMEE}

MMEE primarily targets full dense attention, where operations are regular and static. For static sparse attention \cite{qiu2020blockwise, beltagy2020longformer, zaheer2020big}, computation remains structured, and MMEE remains applicable with a modified performance model. In contrast, dynamic sparse attention \cite{zhu2023biformer, liu2022dynamic} involves data-dependent operations, making accurate dataflow capture and performance estimation challenging for MMEE. This is left for future work.

Beyond its role as a standalone dataflow optimizer, MMEE can be served as a high-level optimization pass for a modern MLIR-based compiler stack \cite{mliraie2023, lattner2021mlir, tillet2019triton, abts2022software, 9895630}. In this workflow, MMEE sits between the high-level dialect, which provides the computational graph, and the low-level backend dialect, which handles code generation. By matrix-form enumeration, MMEE rapidly determines the optimal tiling, computation ordering and buffer management for a given hardware template. 
These parameters can be leveraged in generating highly efficient kernels and code.
This integration allows MMEE to bridge the gap between abstract operator representations and the compiler, navigating complex cross-operator spaces that standard compilers fall short of.

\section{Related Work}

Intra-operator mapping, which optimizes individual operators, has been widely studied \cite{chen2016eyeriss, kao2020gamma, kwon2018maeri, venkatesan2019magnet, xiao2021hasco, farabet2011neuflow, reagen2017case, fu2021auto}. Prior work includes heuristic \cite{chen2019eyeriss, farabet2011neuflow, chen2014diannao, jouppi2017datacenter} and automated approaches \cite{stoutchinin2019optimally, lu2021tenet, liang2022deep, huang2021cosa}. For cross-operator dataflow mapping, early efforts \cite{alwani2016fused, song2017pipelayer} demonstrated CNN fusion. Recent frameworks \cite{kao2022dnnfuser, gao2019tangram, zhou2024pipefuser, zheng2020efficient} extend fusion strategies to general DNNs. Orojenesis \cite{huang2024mind} targets matrix multiplication fusion via exhaustive tiling guided by computation ordering templates. TileFlow \cite{zheng2023tileflow} introduces a tree representation and uses genetic algorithms and Monte Carlo Tree Search to explore tiling, computation ordering, and buffer management. SET \cite{cai2023inter} creates a tree notation emphasizing batch splitting, while Crane \cite{gong2025crane} improves scheduling quality and speed by hierarchical table formats. MAGIS \cite{chen2024magis} coordinated both graph transformation and scheduling to reduce peak memory without increasing latency. FlashAttention \cite{dao2022flashattention, dao2023flashattention} applies the online softmax \cite{milakov2018online} to reduce IO access. FLAT \cite{kao2023flat} exhaustively searches tiling space. FuseMax \cite{nayak2024fusemax} leverages an extended Einsum notation to jointly optimize PE array design and attention dataflow.

\section{Conclusion}

This work proposes \textbf{MMEE}, a framework for efficiently exploring attention fusion dataflows. MMEE systematically captures key design elements and introduces a traversal-free, branch-free analytical performance model. By combining pruning with matrix-based evaluation, MMEE rapidly enumerates various design tradeoffs. Experimental results show that MMEE outperforms state-of-the-art methods in both solution quality and optimization speed.


\bibliographystyle{IEEEtranS}
\bibliography{refs}

@inproceedings{lu2021tenet,
  title={Tenet: A Framework for Modeling Tensor Dataflow Based on Relation-centric Notation},
  author={Lu, Liqiang and Guan, Naiqing and Wang, Yuyue and Jia, Liancheng and Luo, Zizhang and Yin, Jieming and Cong, Jason and Liang, Yun},
  booktitle={International Symposium on Computer Architecture},
  year={2021},
}

@inproceedings{parashar2019timeloop,
  title={Timeloop: A Systematic Approach to {DNN} Accelerator Evaluation},
  author={Parashar, Angshuman and Raina, Priyanka and Shao, Yakun Sophia and Chen, Yu-Hsin and Ying, Victor A and Mukkara, Anurag and Venkatesan, Rangharajan and Khailany, Brucek and Keckler, Stephen W and Emer, Joel},
  booktitle={International Symposium on Performance Analysis of Systems and Software},
  year={2019},
}

@inproceedings{liang2022deep,
  title={Deep Learning Toolkit-accelerated Analytical Co-optimization of {CNN} Hardware and Dataflow},
  author={Liang, Rongjian and Song, Jianfeng and Bo, Yuan and Hu, Jiang},
  booktitle={International Conference on Computer-Aided Design},
  year={2022}
}

@article{stoutchinin2019optimally,
  title={Optimally Scheduling {CNN} Convolutions for Efficient Memory Access},
  author={Stoutchinin, Arthur and Conti, Francesco and Benini, Luca},
  journal={arXiv preprint arXiv:1902.01492},
  year={2019}
}

@inproceedings{zheng2023tileflow,
  title={Tileflow: A Framework for Modeling Fusion Dataflow via Tree-based Analysis},
  author={Zheng, Size and Chen, Siyuan and Gao, Siyuan and Jia, Liancheng and Sun, Guangyu and Wang, Runsheng and Liang, Yun},
  booktitle={International Symposium on Microarchitecture},
  year={2023}
}

@inproceedings{mei2023defines,
  title={Defines: Enabling Fast Exploration of the Depth-First Scheduling Space for {DNN} Accelerators Through Analytical Modeling},
  author={Mei, Linyan and Goetschalckx, Koen and Symons, Arne and Verhelst, Marian},
  booktitle={International Symposium on High-Performance Computer Architecture},
  year={2023},
}

@inproceedings{kao2023flat,
  title={{FLAT:} An Optimized Dataflow for Mitigating Attention Bottlenecks},
  author={Kao, Sheng-Chun and Subramanian, Suvinay and Agrawal, Gaurav and Yazdanbakhsh, Amir and Krishna, Tushar},
  booktitle={International Conference on Architectural Support for Programming Languages and Operating Systems},
  year={2023}
}

@inproceedings{huang2024mind,
  title={{Mind the Gap:} Attainable Data Movement and Operational Intensity Bounds for Tensor Algorithms},
  author={Huang, Qijing and Tsai, Po-An and Emer, Joel S and Parashar, Angshuman},
  booktitle={International Symposium on Computer Architecture},
  year={2024},
}

@inproceedings{yang2020interstellar,
  title={Interstellar: Using Halide’s Scheduling Language to Analyze {DNN} Accelerators},
  author={Yang, Xuan and Gao, Mingyu and Liu, Qiaoyi and Setter, Jeff and Pu, Jing and Nayak, Ankita and Bell, Steven and Cao, Kaidi and Ha, Heonjae and Raina, Priyanka and others},
  booktitle={International Conference on Architectural Support for Programming Languages and Operating Systems},
  year={2020}
}

@inproceedings{wu2019accelergy,
  title={Accelergy: An Architecture-level Energy Estimation Methodology for Accelerator Designs},
  author={Wu, Yannan Nellie and Emer, Joel S and Sze, Vivienne},
  booktitle={International Conference on Computer-Aided Design},
  year={2019},
}

@inproceedings{cai2023inter,
  title={Inter-layer Scheduling Space Definition and Exploration for Tiled Accelerators},
  author={Cai, Jingwei and Wei, Yuchen and Wu, Zuotong and Peng, Sen and Ma, Kaisheng},
  booktitle={International Symposium on Computer Architecture},
  year={2023}
}

@article{dao2022flashattention,
  title={{FlashAttention:} Fast and Memory-efficient Exact Attention with {IO}-Awareness},
  author={Dao, Tri and Fu, Dan and Ermon, Stefano and Rudra, Atri and R{\'e}, Christopher},
  journal={Advances in Neural Information Processing Systems},
  year={2022}
}

@article{dao2023flashattention,
  title={{FlashAttention-2:} Faster Attention with Better Parallelism and Work Partitioning},
  author={Dao, Tri},
  journal={arXiv preprint arXiv:2307.08691},
  year={2023}
}

@inproceedings{zheng2023chimera,
  title={Chimera: An Analytical Optimizing Framework for Effective Compute-intensive Operators Fusion},
  author={Zheng, Size and Chen, Siyuan and Song, Peidi and Chen, Renze and Li, Xiuhong and Yan, Shengen and Lin, Dahua and Leng, Jingwen and Liang, Yun},
  booktitle={International Symposium on High-Performance Computer Architecture},
  year={2023},
}

@article{shao2023efficient,
  title={An Efficient Training Accelerator for Transformers with Hardware-algorithm Co-optimization},
  author={Shao, Haikuo and Lu, Jinming and Wang, Meiqi and Wang, Zhongfeng},
  journal={Transactions on Very Large Scale Integration Systems},
  year={2023},
}

@inproceedings{jouppi2017datacenter,
  title={In-Datacenter Performance Analysis of a Tensor Processing Unit},
  author={Jouppi, Norman P and Young, Cliff and Patil, Nishant and Patterson, David and Agrawal, Gaurav and Bajwa, Raminder and Bates, Sarah and Bhatia, Suresh and Boden, Nan and Borchers, Al and others},
  booktitle={International Symposium on Computer Architecture},
  year={2017}
}

@inproceedings{nayak2024fusemax,
  title={{FuseMax:} Leveraging Extended Einsums to Optimize Attention Accelerator Design},
  author={Nayak, Nandeeka and Wu, Xinrui and Odemuyiwa, Toluwanimi O and Pellauer, Michael and Emer, Joel S and Fletcher, Christopher W},
  booktitle={International Symposium on Microarchitecture},
  year={2024},
}

@article{chen2016eyeriss,
  title={Eyeriss: An Energy-Efficient Reconfigurable Accelerator for Deep Convolutional Neural Networks},
  author={Chen, Yu-Hsin and Krishna, Tushar and Emer, Joel S and Sze, Vivienne},
  journal={Journal of Solid-State Circuits},
  year={2016},
}

@inproceedings{chen2020communication,
  title={Communication Lower Bound in Convolution Accelerators},
  author={Chen, Xiaoming and Han, Yinhe and Wang, Yu},
  booktitle={International Symposium on High Performance Computer Architecture},
  year={2020},
}

@article{cai2022optimus,
  title={Optimus: An Operator Fusion Framework for Deep Neural Networks},
  author={Cai, Xuyi and Wang, Ying and Zhang, Lei},
  journal={Transactions on Embedded Computing Systems},
  year={2022},
}

@inproceedings{kao2020gamma,
  title={{GAMMA:} Automating the Hw Mapping of {DNN} Models on Accelerators via Genetic Algorithm},
  author={Kao, Sheng-Chun and Krishna, Tushar},
  booktitle={International Conference on Computer-Aided Design},
  year={2020}
}

@article{kwon2018maeri,
  title={{MAERI:} Enabling Flexible Dataflow Mapping Over {DNN} Accelerators via Reconfigurable Interconnects},
  author={Kwon, Hyoukjun and Samajdar, Ananda and Krishna, Tushar},
  journal={International Conference on Architectural Support for Programming Languages and Operating Systems},
  year={2018},
}

@inproceedings{venkatesan2019magnet,
  title={{MAGNet:} A Modular Accelerator Generator for Neural Networks},
  author={Venkatesan, Rangharajan and Shao, Yakun Sophia and Wang, Miaorong and Clemons, Jason and Dai, Steve and Fojtik, Matthew and Keller, Ben and Klinefelter, Alicia and Pinckney, Nathaniel and Raina, Priyanka and others},
  booktitle={International Conference on Computer-Aided Design},
  year={2019},
}

@inproceedings{xiao2021hasco,
  title={{HASCO:} Towards Agile Hardware and Software Co-design for Tensor Computation},
  author={Xiao, Qingcheng and Zheng, Size and Wu, Bingzhe and Xu, Pengcheng and Qian, Xuehai and Liang, Yun},
  booktitle={International Symposium on Computer Architecture},
  year={2021},
}

@inproceedings{farabet2011neuflow,
  title={{NeuFlow:} A Runtime Reconfigurable Dataflow Processor for Vision},
  author={Farabet, Cl{\'e}ment and Martini, Berin and Corda, Benoit and Akselrod, Polina and Culurciello, Eugenio and LeCun, Yann},
  booktitle={Computer Vision and Pattern Recognition Workshops},
  year={2011},
}

@inproceedings{reagen2017case,
  title={A Case for Efficient Accelerator Design Space Exploration via Bayesian Optimization},
  author={Reagen, Brandon and Hern{\'a}ndez-Lobato, Jos{\'e} Miguel and Adolf, Robert and Gelbart, Michael and Whatmough, Paul and Wei, Gu-Yeon and Brooks, David},
  booktitle={International Symposium on Low Power Electronics and Design},
  year={2017},
}

@inproceedings{fu2021auto,
  title={{Auto-NBA:} Efficient and Effective Search Over the Joint Space of Networks, Bitwidths, and Accelerators},
  author={Fu, Yonggan and Zhang, Yongan and Zhang, Yang and Cox, David and Lin, Yingyan},
  booktitle={International Conference on Machine Learning},
  year={2021},
}

@inproceedings{alwani2016fused,
  title={Fused-layer {CNN} Accelerators},
  author={Alwani, Manoj and Chen, Han and Ferdman, Michael and Milder, Peter},
  booktitle={International Symposium on Microarchitecture},
  year={2016},
}

@article{kao2022dnnfuser,
  title={{DNNFuser:} Generative Pre-trained Transformer as a Generalized Mapper for Layer Fusion in {DNN} Accelerators},
  author={Kao, Sheng-Chun and Huang, Xiaoyu and Krishna, Tushar},
  journal={arXiv preprint arXiv:2201.11218},
  year={2022}
}

@article{zheng2020efficient,
  title={Efficient Scheduling of Irregular Network Structures on {CNN} Accelerators},
  author={Zheng, Shixuan and Zhang, Xianjue and Ou, Daoli and Tang, Shibin and Liu, Leibo and Wei, Shaojun and Yin, Shouyi},
  journal={Transactions on Computer-Aided Design of Integrated Circuits and Systems},
  year={2020},
}

@inproceedings{gao2019tangram,
  title={{TANGRAM}: Optimized Coarse-grained Dataflow for Scalable {NN} Accelerators},
  author={Gao, Mingyu and Yang, Xuan and Pu, Jing and Horowitz, Mark and Kozyrakis, Christos},
  booktitle={International Conference on Architectural Support for Programming Languages and Operating Systems},
  year={2019}
}

@inproceedings{song2017pipelayer,
  title={PipeLayer: A Pipelined {ReRAM}-Based Accelerator for Deep Learning},
  author={Song, Linghao and Qian, Xuehai and Li, Hai and Chen, Yiran},
  booktitle={International Symposium on High Performance Computer Architecture},
  year={2017},
}

@inproceedings{zhou2024pipefuser,
  title={{PipeFuser}: Building Flexible Pipeline Architecture for {DNN} Accelerators via Layer Fusion},
  author={Zhou, Xilang and Li, Shuyang and Lu, Haodong and Wang, Kun},
  booktitle={Asia and South Pacific Design Automation Conference},
  year={2024},
}

@article{chen2019eyeriss,
  title={Eyeriss v2: A Flexible Accelerator for Emerging Deep Neural Networks on Mobile Devices},
  author={Chen, Yu-Hsin and Yang, Tien-Ju and Emer, Joel and Sze, Vivienne},
  journal={Journal on Emerging and Selected Topics in Circuits and Systems},
  year={2019},
}

@inproceedings{chang2023multifuse,
  title={MultiFuse: Efficient Cross Layer Fusion for {DNN} Accelerators With Multi-Level Memory Hierarchy},
  author={Chang, Chia-Wei and Liou, Jing-Jia and Huang, Chih-Tsun and Hsu, Wei-Chung and Lu, Juin-Ming},
  booktitle={International Conference on Computer Design},
  year={2023},
}

@inproceedings{zhou2024ml,
  title={{ML-Fusion}: Determining Memory Levels for Data Reuse Between {DNN} Layers},
  author={Zhou, Zikang and Duan, Xuyang and Chen, Kaiqi and Chen, Yaqi and Han, Jun},
  booktitle={Proceedings of the Great Lakes Symposium on {VLSI}},
  year={2024}
}

@article{conneau2019cross,
  title={Cross-Lingual Language Model Pretraining},
  author={Conneau, Alexis and Lample, Guillaume},
  journal={Advances in Neural Information Processing Systems},
  year={2019}
}

@article{raffel2020exploring,
  title={Exploring the Limits of Transfer Learning With a Unified Text-To-Text Transformer},
  author={Raffel, Colin and Shazeer, Noam and Roberts, Adam and Lee, Katherine and Narang, Sharan and Matena, Michael and Zhou, Yanqi and Li, Wei and Liu, Peter J},
  journal={Journal of Machine Learning Research},
  year={2020}
}

@article{dosovitskiy2020image,
  title={An Image Is Worth 16x16 Words: Transformers for Image Recognition at Scale},
  author={Dosovitskiy, Alexey and Beyer, Lucas and Kolesnikov, Alexander and Weissenborn, Dirk and Zhai, Xiaohua and Unterthiner, Thomas and Dehghani, Mostafa and Minderer, Matthias and Heigold, Georg and Gelly, Sylvain and others},
  journal={arXiv preprint arXiv:2010.11929},
  year={2020}
}

@inproceedings{lin2022cat,
  title={{CAT}: Cross Attention in Vision Transformer},
  author={Lin, Hezheng and Cheng, Xing and Wu, Xiangyu and Shen, Dong},
  booktitle={International Conference on Multimedia and Expo},
  year={2022},
}

@inproceedings{zhang2022styleswin,
  title={{StyleSwin}: Transformer-Based {GAN} for High-Resolution Image Generation},
  author={Zhang, Bowen and Gu, Shuyang and Zhang, Bo and Bao, Jianmin and Chen, Dong and Wen, Fang and Wang, Yong and Guo, Baining},
  booktitle={Conference on Computer Vision and Pattern Recognition},
  year={2022}
}

@article{naveen2021transformer,
  title={Transformer Models for Enhancing {AttnGAN} Based Text to Image Generation},
  author={Naveen, S and Kiran, MS S Ram and Indupriya, M and Manikanta, TV and Sudeep, PV},
  journal={Image and Vision Computing},
  year={2021},
}

@inproceedings{shen2021efficient,
  title={Efficient Attention: Attention With Linear Complexities},
  author={Shen, Zhuoran and Zhang, Mingyuan and Zhao, Haiyu and Yi, Shuai and Li, Hongsheng},
  booktitle={Conference on Applications of Computer Vision},
  year={2021}
}

@inproceedings{keles2023computational,
  title={On the Computational Complexity of Self-Attention},
  author={Keles, Feyza Duman and Wijewardena, Pruthuvi Mahesakya and Hegde, Chinmay},
  booktitle={International Conference on Algorithmic Learning Theory},
  year={2023},
}

@inproceedings{devlin2019bert,
  title={{BERT}: Pre-training of Deep Bidirectional Transformers for Language Understanding},
  author={Devlin, Jacob and Chang, Ming-Wei and Lee, Kenton and Toutanova, Kristina},
  booktitle={Conference of the North {American} Chapter of the Association for Computational Linguistics: Human Language Technologies},
  year={2019}
}

@article{brown2020language,
  title={Language Models are Few-shot Learners},
  author={Tom B. Brown and Benjamin Mann and Nick Ryder and Melanie Subbiah and Jared Kaplan and Prafulla Dhariwal and Arvind Neelakantan and Pranav Shyam and Girish Sastry and Amanda Askell and Sandhini Agarwal and Ariel Herbert-Voss and Gretchen Krueger and Tom Henighan and Rewon Child and Aditya Ramesh and Daniel M. Ziegler and Jeffrey Wu and Clemens Winter and Christopher Hesse and Mark Chen and Eric Sigler and Mateusz Litwin and Scott Gray and Benjamin Chess and Jack Clark and Christopher Berner and Sam McCandlish and Alec Radford and Ilya Sutskever and Dario Amodei},
  journal={Advances in Neural Information Processing Systems},
  year={2020}
}

@article{chowdhery2023palm,
  title={PaLM: Scaling Language Modeling with Pathways},
  author={Aakanksha Chowdhery and Sharan Narang and Jacob Devlin and Maarten Bosma and Gaurav Mishra and Adam Roberts and Paul Barham and Hyung Won Chung and Charles Sutton and Sebastian Gehrmann and Parker Schuh and Kensen Shi and Sasha Tsvyashchenko and Joshua Maynez and Abhishek Rao and Parker Barnes and Yi Tay and Noam Shazeer and Vinodkumar Prabhakaran and Emily Reif and Nan Du and Ben Hutchinson and Reiner Pope and James Bradbury and Jacob Austin and Michael Isard and Guy Gur-Ari and Pengcheng Yin and Toju Duke and Anselm Levskaya and Sanjay Ghemawat and Sunipa Dev and Henryk Michalewski and Xavier Garcia and Vedant Misra and Kevin Robinson and Liam Fedus and Denny Zhou and Daphne Ippolito and David Luan and Hyeontaek Lim and Barret Zoph and Alexander Spiridonov and Ryan Sepassi and David Dohan and Shivani Agrawal and Mark Omernick and Andrew M. Dai and Thanumalayan Sankaranarayana Pillai and Marie Pellat and Aitor Lewkowycz and Erica Moreira and Rewon Child and Oleksandr Polozov and Katherine Lee and Zongwei Zhou and Xuezhi Wang and Brennan Saeta and Mark Diaz and Orhan Firat and Michele Catasta and Jason Wei and Kathy Meier-Hellstern and Douglas Eck and Jeff Dean and Slav Petrov and Noah Fiedel},
  journal={Journal of Machine Learning Research},
  year={2023}
}

@article{shazeer2019fast,
  title={Fast Transformer Decoding: {One} Write-head is All You Need},
  author={Shazeer, Noam},
  journal={arXiv preprint arXiv:1911.02150},
  year={2019}
}

@inproceedings{ainslie2023gqa,
  title={GQA: Training Generalized Multi-query Transformer Models from Multi-head Checkpoints},
  author={Ainslie, Joshua and Lee-Thorp, James and De Jong, Michiel and Zemlyanskiy, Yury and Lebr{\'o}n, Federico and Sanghai, Sumit},
  booktitle={Proceedings of the 2023 Conference on Empirical Methods in Natural Language Processing},
  pages={4895--4901},
  year={2023}
}

@inproceedings{wang2021spatten,
  title={Spatten: Efficient Sparse Attention Architecture with Cascade Token and Head Pruning},
  author={Wang, Hanrui and Zhang, Zhekai and Han, Song},
  booktitle={International Symposium on High-Performance Computer Architecture},
  year={2021},
}

@article{tay2020long,
  title={Long Range Arena: A Benchmark for Efficient Transformers},
  author={Tay, Yi and Dehghani, Mostafa and Abnar, Samira and Shen, Yikang and Bahri, Dara and Pham, Philip and Rao, Jinfeng and Yang, Liu and Ruder, Sebastian and Metzler, Donald},
  journal={arXiv preprint arXiv:2011.04006},
  year={2020}
}

@article{beltagy2020longformer,
  title={Longformer: The Long-document Transformer},
  author={Beltagy, Iz and Peters, Matthew E and Cohan, Arman},
  journal={arXiv preprint arXiv:2004.05150},
  year={2020}
}

@article{kitaev2020reformer,
  title={Reformer: The Efficient Transformer},
  author={Kitaev, Nikita and Kaiser, {\L}ukasz and Levskaya, Anselm},
  journal={arXiv preprint arXiv:2001.04451},
  year={2020}
}

@article{desislavov2021compute,
  title={Compute and Energy Consumption Trends in Deep Learning Inference},
  author={Desislavov, Radosvet and Mart{\'\i}nez-Plumed, Fernando and Hern{\'a}ndez-Orallo, Jos{\'e}},
  journal={arXiv preprint arXiv:2109.05472},
  year={2021}
}

@inproceedings{zadeh2020gobo,
  title={Gobo: Quantizing Attention-based {NLP} Models for Low Latency and Energy Efficient Inference},
  author={Zadeh, Ali Hadi and Edo, Isak and Awad, Omar Mohamed and Moshovos, Andreas},
  booktitle={International Symposium on Microarchitecture},
  year={2020},
}

@article{chen2014diannao,
  title={Diannao: A Small-footprint High-throughput Accelerator for Ubiquitous Machine-learning},
  author={Chen, Tianshi and Du, Zidong and Sun, Ninghui and Wang, Jia and Wu, Chengyong and Chen, Yunji and Temam, Olivier},
  journal={SIGARCH Computer Architecture News},
  year={2014},
}

@article{milakov2018online,
  title={Online Normalizer Calculation for Softmax},
  author={Milakov, Maxim and Gimelshein, Natalia},
  journal={arXiv preprint arXiv:1805.02867},
  year={2018}
}

@misc{nvdla2017,
  author       = {{NVIDIA}},
  title        = {{NVDLA Deep Learning Accelerator}},
  howpublished = {\url{http://nvdla.org}},
  year         = {2017},
  note         = {Accessed: 2025-06-16}
}

@misc{coral_intro_guide,
  author       = {{Google}},
  title        = {{Coral NPU}},
  howpublished = {\url{https://developers.google.com/coral/guides/intro}},
  year         = {2025},
  note         = {Accessed: 2025-10-16}
}

@inproceedings{seshadri2022evaluation,
  title={An Evaluation of Edge TPU Accelerators for Convolutional Neural Networks},
  author={Seshadri, Kiran and Akin, Berkin and Laudon, James and Narayanaswami, Ravi and Yazdanbakhsh, Amir},
  booktitle={2022 IEEE International Symposium on Workload Characterization (IISWC)},
  pages={79--91},
  year={2022},
  organization={IEEE}
}

@misc{tillet2020triton_tutorials,
  author       = {{Philippe Tillet}},
  title        = {{Triton Getting Started Tutorials}},
  howpublished = {\url{https://triton‑lang.org/main/getting‑started/tutorials/index.html}},
  year         = {2020},
  note         = {Accessed: 2025‑07‑27}
}

@inproceedings{gong2025crane,
  title={Crane: Inter-Layer Scheduling Framework for DNN Inference and Training Co-Support on Tiled Architecture},
  author={Gong, Yu and Huang, Lingyi and Chang, Haodong and Liang, Rongjian and Yang, Cheng and Tang, Zhexiang and Hu, Jiang and Yuan, Bo},
  booktitle={Proceedings of the 58th IEEE/ACM International Symposium on Microarchitecture{\textregistered}},
  pages={1250--1263},
  year={2025}
}

@inproceedings{zhao2020dnn,
  title={DNN-Chip Predictor: An Analytical Performance Predictor for DNN Accelerators with Various Dataflows and Hardware Architectures},
  author={Zhao, Yang and Li, Chaojian and Wang, Yue and Xu, Pengfei and Zhang, Yongan and Lin, Yingyan},
  booktitle={ICASSP 2020-2020 IEEE International Conference on Acoustics, Speech and Signal Processing (ICASSP)},
  pages={1593--1597},
  year={2020},
  organization={IEEE}
}

@inproceedings{huang2021cosa,
  title={Cosa: Scheduling by Constrained Optimization for Spatial Accelerators},
  author={Huang, Qijing and Kang, Minwoo and Dinh, Grace and Norell, Thomas and Kalaiah, Aravind and Demmel, James and Wawrzynek, John and Shao, Yakun Sophia},
  booktitle={2021 ACM/IEEE 48th Annual International Symposium on Computer Architecture (ISCA)},
  pages={554--566},
  year={2021},
  organization={IEEE}
}

@inproceedings{wei2020design,
  title={Design Space Exploration for Softmax Implementations},
  author={Wei, Zhigang and Arora, Aman and Patel, Pragenesh and John, Lizy},
  booktitle={2020 IEEE 31st International Conference on Application-specific Systems, Architectures and Processors (ASAP)},
  pages={45--52},
  year={2020},
  organization={IEEE}
}

@article{lin2025systolicattention,
  title={SystolicAttention: Fusing FlashAttention within a Single Systolic Array},
  author={Lin, Jiawei and Li, Yuanlong and Chen, Guokai and Bourgeat, Thomas},
  journal={arXiv preprint arXiv:2507.11331},
  year={2025}
}

@article{shah2024flashattention,
  title={FlashAttention-3: Fast and Accurate Attention with Asynchrony and Low-precision},
  author={Shah, Jay and Bikshandi, Ganesh and Zhang, Ying and Thakkar, Vijay and Ramani, Pradeep and Dao, Tri},
  journal={Advances in Neural Information Processing Systems},
  volume={37},
  pages={68658--68685},
  year={2024}
}

@article{pan2025fasttree,
  title={FastTree: Optimizing Attention Kernel and Runtime for Tree-Structured LLM Inference},
  author={Pan, Zaifeng and Ding, Yitong and Guan, Yue and Wang, Zheng and Yu, Zhongkai and Tang, Xulong and Wang, Yida and Ding, Yufei},
  journal={Proceedings of Machine Learning and Systems},
  volume={7},
  year={2025}
}

@article{ho2024block,
  title={Block Transformer: Global-to-Local Language Modeling for Fast Inference},
  author={Ho, Namgyu and Bae, Sangmin and Kim, Taehyeon and Jo, Hyunjik and Kim, Yireun and Schuster, Tal and Fisch, Adam and Thorne, James and Yun, Se-Young},
  journal={Advances in Neural Information Processing Systems},
  volume={37},
  pages={48740--48783},
  year={2024}
}

@inproceedings{kwon2023efficient,
  title={Efficient Memory Management for Large Language Model Serving with PagedAttention},
  author={Kwon, Woosuk and Li, Zhuohan and Zhuang, Siyuan and Sheng, Ying and Zheng, Lianmin and Yu, Cody Hao and Gonzalez, Joseph and Zhang, Hao and Stoica, Ion},
  booktitle={Proceedings of the 29th symposium on operating systems principles},
  pages={611--626},
  year={2023}
}

@inproceedings{li2025tritonbench,
  title={TritonBench: Benchmarking Large Language Model Capabilities for Generating Triton Operators},
  author={Li, Jianling and Li, Shangzhan and Gao, Zhenye and Shi, Qi and Li, Yuxuan and Wang, Zefan and Huang, Jiacheng and WangHaojie, WangHaojie and Wang, Jianrong and Han, Xu and others},
  booktitle={Findings of the Association for Computational Linguistics: ACL 2025},
  pages={23053--23066},
  year={2025}
}

@inproceedings{tillet2019triton,
  title={Triton: An Intermediate Language and Compiler for Tiled Neural Network Computations},
  author={Tillet, Philippe and Kung, Hsiang-Tsung and Cox, David},
  booktitle={Proceedings of the 3rd ACM SIGPLAN International Workshop on Machine Learning and Programming Languages},
  pages={10--19},
  year={2019}
}

@article{li2025tl,
  title={TL: Automatic End-to-End Compiler of Tile-Based Languages for Spatial Dataflow Architectures},
  author={Li, Wei and Bai, Zhenyu and Wang, Heru and Dangi, Pranav and Zhang, Zhiqiang and Tan, Cheng and Lan, Huiying and Wong, Weng-Fai and Mitra, Tulika},
  journal={arXiv preprint arXiv:2512.22168},
  year={2025}
}

@misc{aws_neuroncore_v2,
  author       = {{AWS Neuron Team}},
  title        = {NeuronCore-v2 Architecture},
  howpublished = {\url{https://awsdocs-neuron.readthedocs-hosted.com/en/latest/about-neuron/arch/neuron-hardware/neuron-core-v2.html}},
  year         = {2024},
  note         = {Accessed: 2026-02-22}
}

@inproceedings{sheng2023flexgen,
  title={Flexgen: High-Throughput Generative Inference of Large Language Models with A Single GPU},
  author={Sheng, Ying and Zheng, Lianmin and Yuan, Binhang and Li, Zhuohan and Ryabinin, Max and Chen, Beidi and Liang, Percy and R{\'e}, Christopher and Stoica, Ion and Zhang, Ce},
  booktitle={International Conference on Machine Learning},
  pages={31094--31116},
  year={2023},
  organization={PMLR}
}

@article{zhang2023h2o,
  title={H2O: Heavy-Hitter Oracle for Efficient Generative Inference of Large Language Models},
  author={Zhang, Zhenyu and Sheng, Ying and Zhou, Tianyi and Chen, Tianlong and Zheng, Lianmin and Cai, Ruisi and Song, Zhao and Tian, Yuandong and R{\'e}, Christopher and Barrett, Clark and others},
  journal={Advances in Neural Information Processing Systems},
  volume={36},
  pages={34661--34710},
  year={2023}
}

@article{liu2024kivi,
  title={KIVI: A Tuning-Free Asymmetric 2bit Quantization for KV Cache},
  author={Liu, Zirui and Yuan, Jiayi and Jin, Hongye and Zhong, Shaochen and Xu, Zhaozhuo and Braverman, Vladimir and Chen, Beidi and Hu, Xia},
  journal={arXiv preprint arXiv:2402.02750},
  year={2024}
}

@inproceedings{zhong2024distserve,
  title={DistServe: Disaggregating Prefill and Decoding for Goodput-Optimized Large Language Model Serving},
  author={Zhong, Yinmin and Liu, Shengyu and Chen, Junda and Hu, Jianbo and Zhu, Yibo and Liu, Xuanzhe and Jin, Xin and Zhang, Hao},
  booktitle={18th USENIX Symposium on Operating Systems Design and Implementation (OSDI 24)},
  pages={193--210},
  year={2024}
}

@article{zhang2025spad,
  title={SPAD: Specialized Prefill and Decode Hardware for Disaggregated LLM Inference},
  author={Zhang, Hengrui and Patel, Pratyush and Ning, August and Wentzlaff, David},
  journal={arXiv preprint arXiv:2510.08544},
  year={2025}
}

@inproceedings{feng2024long,
  title={Long Sequence Modeling with Attention Tensorization: From Sequence to Tensor Learning},
  author={Feng, Aosong and Ying, Rex and Tassiulas, Leandros},
  booktitle={Findings of the Association for Computational Linguistics: EMNLP 2024},
  pages={14642--14655},
  year={2024}
}

@article{stallone2024scaling,
  title={Scaling Granite Code Models to 128K Context},
  author={Stallone, Matt and Saxena, Vaibhav and Karlinsky, Leonid and McGinn, Bridget and Bula, Tim and Mishra, Mayank and Soria, Adriana Meza and Zhang, Gaoyuan and Prasad, Aditya and Shen, Yikang and others},
  journal={arXiv preprint arXiv:2407.13739},
  year={2024}
}

@inproceedings{qiu2020blockwise,
  title={Blockwise Self-Attention for Long Document Understanding},
  author={Qiu, Jiezhong and Ma, Hao and Levy, Omer and Yih, Wen-tau and Wang, Sinong and Tang, Jie},
  booktitle={Findings of the Association for Computational Linguistics: EMNLP 2020},
  pages={2555--2565},
  year={2020}
}

@article{zaheer2020big,
  title={Big Bird: Transformers for Longer Sequences},
  author={Zaheer, Manzil and Guruganesh, Guru and Dubey, Kumar Avinava and Ainslie, Joshua and Alberti, Chris and Ontanon, Santiago and Pham, Philip and Ravula, Anirudh and Wang, Qifan and Yang, Li and others},
  journal={Advances in neural information processing systems},
  volume={33},
  pages={17283--17297},
  year={2020}
}

@inproceedings{zhu2023biformer,
  title={BiFormer: Vision Transformer with Bi-Level Routing Attention},
  author={Zhu, Lei and Wang, Xinjiang and Ke, Zhanghan and Zhang, Wayne and Lau, Rynson WH},
  booktitle={Proceedings of the IEEE/CVF conference on computer vision and pattern recognition},
  pages={10323--10333},
  year={2023}
}

@article{liu2022dynamic,
  title={Dynamic Sparse Attention for Scalable Transformer Acceleration},
  author={Liu, Liu and Qu, Zheng and Chen, Zhaodong and Tu, Fengbin and Ding, Yufei and Xie, Yuan},
  journal={IEEE Transactions on Computers},
  volume={71},
  number={12},
  pages={3165--3178},
  year={2022},
  publisher={IEEE}
}

@inproceedings{chen2024magis,
  title={MAGIS: Memory Optimization via Coordinated Graph Transformation and Scheduling for DNN},
  author={Chen, Renze and Ding, Zijian and Zheng, Size and Zhang, Chengrui and Leng, Jingwen and Liu, Xuanzhe and Liang, Yun},
  booktitle={Proceedings of the 29th ACM International Conference on Architectural Support for Programming Languages and Operating Systems, Volume 3},
  pages={607--621},
  year={2024}
}

@article{zheng2024sglang,
  title={SGLang: Efficient Execution of Structured Language Model Programs},
  author={Zheng, Lianmin and Yin, Liangsheng and Xie, Zhiqiang and Sun, Chuyue Livia and Huang, Jeff and Yu, Cody Hao and Cao, Shiyi and Kozyrakis, Christos and Stoica, Ion and Gonzalez, Joseph E and others},
  journal={Advances in neural information processing systems},
  volume={37},
  pages={62557--62583},
  year={2024}
}

@article{lee2024resa,
  title={ReSA: Reconfigurable Systolic Array for Multiple Tiny DNN Tensors},
  author={Lee, Ching-Jui and Yeh, Tsung Tai},
  journal={ACM Transactions on Architecture and Code Optimization},
  volume={21},
  number={3},
  pages={1--24},
  year={2024},
  publisher={ACM New York, NY}
}

@article{han2024redas,
  title={ReDas: A Lightweight Architecture for Supporting Fine-Grained Reshaping and Multiple Dataflows on Systolic Array},
  author={Han, Meng and Wang, Liang and Xiao, Limin and Cai, Tianhao and Wang, Zeyu and Xu, Xiangrong and Zhang, Chenhao},
  journal={IEEE Transactions on Computers},
  volume={73},
  number={8},
  pages={1997--2011},
  year={2024},
  publisher={IEEE}
}

@article{wang2022novel,
  title={A Novel Systolic Array Processor with Dynamic Dataflows},
  author={Wang, Bo and Ma, Sheng and Zhu, Guoyi and Yi, Xiao and Xu, Rui},
  journal={Integration},
  volume={85},
  pages={42--47},
  year={2022},
  publisher={Elsevier}
}

@article{wang2025desa,
  title={DESA: Dataflow Efficient Systolic Array for Acceleration of Transformers},
  author={Wang, Zhican and Fan, Hongxiang and He, Guanghui},
  journal={IEEE Transactions on Computers},
  year={2025},
  publisher={IEEE}
}

@misc{mliraie2023,
  author = {{Advanced Micro Devices, Inc.}},
  title = {{MLIR-AIE: MLIR-based Toolchain for AI Engine-enabled Devices}},
  howpublished = {\url{https://xilinx.github.io/mlir-aie/}},
  year = {2023},
  note = {Accessed: 2024-05-22}
}

@inproceedings{lattner2021mlir,
  title={MLIR: Scaling Compiler Infrastructure for Domain Specific Computation},
  author={Lattner, Chris and Amini, Mehdi and Bondhugula, Uday and Cohen, Albert and Davis, Andy and Pienaar, Jacques and Riddle, River and Shpeisman, Tatiana and Vasilache, Nicolas and Zinenko, Oleksandr},
  booktitle={2021 IEEE/ACM International Symposium on Code Generation and Optimization (CGO)},
  pages={2--14},
  year={2021},
  organization={IEEE}
}

@inproceedings{abts2022software,
  title={A Software-defined Tensor Streaming Multiprocessor for Large-scale Machine Learning},
  author={Abts, Dennis and Kimmell, Garrin and Ling, Andrew and Kim, John and Boyd, Matt and Bitar, Andrew and Parmar, Sahil and Ahmed, Ibrahim and DiCecco, Roberto and Han, David and others},
  booktitle={Proceedings of the 49th Annual International Symposium on Computer Architecture (ISCA)},
  pages={567--580},
  year={2022}
}

@INPROCEEDINGS{9895630,
  author={Abts, Dennis and Kim, John and Kimmell, Garrin and Boyd, Matthew and Kang, Kris and Parmar, Sahil and Ling, Andrew and Bitar, Andrew and Ahmed, Ibrahim and Ross, Jonathan},
  booktitle={2022 IEEE Hot Chips 34 Symposium (HCS)}, 
  title={The Groq Software-defined Scale-out Tensor Streaming Multiprocessor : From chips-to-systems architectural overview}, 
  year={2022},
  volume={},
  number={},
  pages={1-69},
  doi={10.1109/HCS55958.2022.9895630}}

\end{document}